\documentclass[twocolumn,lineno,dvipsnames, usenames]{aastex63}
\usepackage[utf8]{inputenc}
\usepackage{amssymb,amsmath,verbatim,mathtools,needspace,enumitem,etoolbox,graphicx,physics,microtype,ulem, breqn}
\usepackage{xspace} 
\usepackage{booktabs}
\usepackage{upgreek} 
\newcommand{\um}{$\upmu$m\xspace}

\newcommand{\LIGOlabMIT}{\affiliation{LIGO Laboratory, Massachusetts Institute of Technology, 185 Albany St, Cambridge, MA 02139, USA}}
\newcommand{\MKI}{\affiliation{Department of Physics and Kavli Institute for Astrophysics and Space Research, Massachusetts Institute of Technology, 77 Massachusetts Ave, Cambridge, MA 02139, USA}}
\newcommand{\Caltech}{\affiliation{Division of Physics, Mathematics, and Astronomy, California Institute of Technology, Pasadena, CA 91125, USA}}
\newcommand{\Paris}{\affiliation{Universit\'{e} Paris-Saclay, CNRS/IN2P3, IJCLab, 91405 Orsay, France}}

\newcommand{\pycbc}{\texttt{PyCBC Live}\xspace}
\newcommand{\bayestar}{\texttt{BAYESTAR}\xspace}
\newcommand{\bilby}{\texttt{bilby}\xspace}
\newcommand{\lalinference}{\texttt{LALInference}\xspace}
\newcommand{\gwemopt}{\texttt{gwemopt}\xspace}
\newcommand{\gwemlightcurves}{\texttt{gwemlightcurves}\xspace}

\begin{document}

\title{An Infrared Search for Kilonovae with the WINTER Telescope. I. Binary Neutron Star Mergers}

\correspondingauthor{Danielle Frostig, Sylvia Biscoveanu}
 \email{frostig@mit.edu, sbisco@mit.edu}
\author[0000-0002-7197-9004]{Danielle Frostig} \MKI
\author[0000-0001-7616-7366]{Sylvia Biscoveanu} \MKI \LIGOlabMIT
\author[0000-0001-6331-112X]{Geoffrey Mo} \MKI \LIGOlabMIT
\author[0000-0003-2758-159X]{Viraj Karambelkar} \Caltech
\author[0000-0001-5078-9044]{Tito Dal Canton} \Paris
\author[0000-0001-5403-3762]{Hsin-Yu Chen} \MKI \LIGOlabMIT
\author{Mansi Kasliwal} \Caltech
\author{Erik Katsavounidis}\MKI \LIGOlabMIT
\author{Nathan P. Lourie}\MKI
\author{Robert A. Simcoe}\MKI
\author[0000-0003-2700-0767]{Salvatore Vitale} \MKI \LIGOlabMIT

\begin{abstract}
The Wide-Field Infrared Transient Explorer (WINTER) is a new 1 $\text{deg}^2$ seeing-limited time-domain survey instrument designed for dedicated near-infrared follow-up of kilonovae from binary neutron star (BNS) and neutron star-black hole mergers. WINTER will observe in the near-infrared Y, J, and short-H bands (0.9-1.7 microns, to $\text{J}_{AB}=21$ magnitudes) on a dedicated 1-meter telescope at Palomar Observatory. To date, most prompt kilonova follow-up has been in optical wavelengths; however, near-infrared emission fades more slowly and depends less on geometry and viewing angle than optical emission. We present an end-to-end simulation of a follow-up campaign during the fourth observing run (O4) of the LIGO, Virgo, and KAGRA interferometers, including simulating 625 BNS mergers, their detection in gravitational waves, low-latency and full parameter estimation skymaps, and a suite of kilonova lightcurves from two different model grids. We predict up to five new kilonovae independently discovered by WINTER during O4, given a realistic BNS merger rate. Using a larger grid of kilonova parameters, we find that kilonova emission is $\approx$2 times longer-lived and red kilonovae are detected $\approx$1.5 times further in the infrared than in the optical. For 90\% localization areas smaller than 150 (450) $\rm{deg}^{2}$, WINTER will be sensitive to more than 10\% of the kilonova model grid out to 350 (200) Mpc. We develop a generalized toolkit to create an optimal BNS follow-up strategy with any electromagnetic telescope and present WINTER's observing strategy with this framework. This toolkit, all simulated gravitational-wave events, and skymaps are made available for use by the community.
\end{abstract}


\section{Introduction}
\label{sec:intro} 
During the second Advanced LIGO-Virgo observing run (O2; \cite{TheLIGOScientific:2014jea,TheVirgo:2014hva}), the binary neutron star (BNS) merger GW170817 led to the first coincident detection of gravitational waves (GWs) with electromagnetic (EM) radiation spanning X-ray to radio frequencies \citep{TheLIGOScientific:2017qsa, GBM:2017lvd}. GW170817 marked the first direct evidence of a kilonova---a thermal transient powered by rapid neutron capture (r-process) nucleosynthesis in the neutron-rich ejecta of BNS mergers~\citep{LIGOScientific:2017vwq, Barnes:2016umi, Coulter:2017wya, Barnes:2013wka, Evans:2017mmy, Goldstein:2017mmi,  Grossman:2013lqa, Haggard:2017qne, Hallinan:2017woc, Kasen:2013xka, Kasliwal:2018fwk, Li:1998bw, Margutti:2017cjl, Metzger:2010sy, Metzger:2019zeh,  Roberts:2011xz, Rosswog:2005su, Tanaka:2013ana, Troja:2017nqp}. Kilonovae offer a unique laboratory to study the production of heavy elements via the r-process, probe the neutron star equation of state~\citep{Margalit:2017dij, Coughlin:2018fis, Coughlin:2018miv, Breschi:2021tbm}, and provide a new class of standard sirens for resolving the Hubble tension \citep{Holz:2005df, Abbott:2017xzu, Chen:2017rfc, Coughlin:2019vtv, Dietrich:2020efo}. The combination of GW alerts and electromagnetic follow-up provides a new tool to expand the small dataset of known kilonovae from BNS (although see \cite{Tanvir:2013pia, Troja:2018ybt, Fong:2020cej}) and neutron star-black hole (NSBH) mergers~\citep{LIGOScientific:2021qlt}.

Despite the successful observation of GW170817 in O2, no new, confirmed electromagnetic counterparts to BNS or NSBH mergers were discovered during the third Advanced LIGO-Virgo observing run (O3; \cite{Abbott:2020niy, LIGOScientific:2021qlt, LIGOScientific:2021usb, Coughlin:2020fwx, Kasliwal:2020wmy, Zhu:2021ysz}).  Many ultraviolet/visible/infrared (UVOIR) teams undertook follow-up observations of the O3 LIGO alerts, including the ASAS-SN \citep{Kochanek:2017wud}, ATLAS \citep{Tonry2018abc}, BOOTES \citep{2021RMxAC..53...75H}, DDOTI \citep{2021MNRAS.507.1401B}, DES-GW \citep{DES:2017kbs}, ENGRAVE \citep{Levan:2020BP}, GOTO \citep{Gompertz:2020cur}, GRANDMA \citep{Antier:2020nuy}, GROWTH \citep{Kasliwal:2020wmy}, MASTER-Net \citep{Lipunov:2017dwd}, SAGUARO;  \citep{Paterson:2020mmd}, Swift UVOT \citep{Oates:2021eyk}, and VINROUGE \citep{Ackley:2020qkz} teams. In addition to searches for UVOIR kilonovae, the Fermi and Swift \citep{Page:2020tnx} spacecraft conducted gamma-ray and X-ray follow-up, respectively, and the Australian Square Kilometre Array Pathfinder searched in radio wavelengths \citep{Dobie:2019ctw}. During O3, spurious GW alerts and alert retractions complicated electromagnetic follow-up strategies \citep{Coughlin:2019zqi}. Additionally, models predict many of the BNS mergers detected in GWs in O3 produced kilonovae that were too faint to be detected by the UVOIR facilities conducting follow-up at the time \citep{Kasliwal:2020wmy, Zhu:2021ysz}. 

Optical-wavelength searches dominated the UVOIR O3 GW follow-up landscape, with dozens of optical telescopes across the globe and in orbit triggered during O3 for GW follow-up. Theoretical models and observational constraints from O3 nondetections predict that the blue, optical emission from a kilonova is angle-dependent, fades rapidly ($<$1 week), and may not be present in all BNS or NSBH mergers \citep{Metzger:2019zeh, Kasen:2013xka, Kasen:2017sxr, Barnes:2016umi, Kasliwal:2020wmy}. In contrast, the near-infrared emission is expected to be isotropic, long-lived ($>$1 week), and ubiquitous in models regardless of mass ratio, viewing angle, or remnant lifetime \citep{Kasen:2017sxr}. Models predict the detection rates of kilonovae in the near-infrared could be up to $\sim8-10$ times higher than in optical wave bands \citep{Zhu:2020ffa}. 

However, due to the high cost-per-pixel of detectors and bright sky backgrounds, the dynamic infrared sky remains largely underexplored compared to optical wavelengths. Existing time-domain, infrared surveys are either restricted to small areas on sky (e.g., the VISTA Variables in the Via Lactea survey covering 520 $\text{deg}^2$ \citep{catelan2011vista} or the UKIRT Deep Extragalactic Survey covering 35 $\text{deg}^2$ \citep{Lawrence:2006de}) or are relatively shallow (e.g., Palomar Gattini IR, $\text{J}\sim16$ mag; \cite{De:2019xhw}). The lack of deep, all-sky reference images limits the number of GW events that can be followed up in the infrared.

The Wide-field Infrared Transient Explorer (WINTER) is a new instrument that will perform the first near-infrared, all-sky survey to $\text{J}_{AB} = 21$ magnitudes and is specially built for GW follow-up \citep{Frostig:2020, Lourie:2020}. WINTER will operate on a dedicated 1-meter telescope at Palomar Observatory that was commissioned in June 2021 with an optical camera on one port, with the infrared WINTER instrument to be added on the second port in late 2021. WINTER's wide, 1 $\text{deg}^2$ field of view can quickly tile the median expected fourth observing run (O4) 33 $\text{deg}^2$ BNS localization contour \citep{Abbott:2020qfu} with rapid-response robotic observing. With three near-infrared filters in the Y, J, and a shortened-H bands (centered at 1.0, 1.2, and 1.6 \um, respectively), WINTER is designed to discover kilonovae and observe them for two weeks or more \citep{Frostig:2020}. 

As an alternative to the traditional, but expensive, mercury-cadmium-telluride (HgCdTe) detectors that dominate the near-infrared landscape, WINTER employs cheaper indium-gallium-arsenide (InGaAs) detectors new to astronomical instrumentation. A prototype instrument confirmed InGaAs detectors achieve background-limited near-infrared photometry without cryogenic cooling \citep{Simcoe:2019aps}, which is required for HgCdTe detectors. WINTER will head a new class of InGaAs-based near-infrared astronomical instruments coming online in the next decade, including the DREAMS telescope in the Southern Hemisphere \citep{DREAMS}. The upcoming HgCdTe-based PRime-focus Infrared Microlensing Experiment telescope will also join WINTER and DREAMS in an upcoming effort to deepen our understanding of the near-infrared time-domain sky.

In this paper, we present an end-to-end simulation of WINTER's performance in O4 and make a case for infrared follow-up of BNS GW signals.
Existing simulations such as \cite{Aasi:2013wya} also present predictions for BNS merger rates detected by the global GW network in O4, but do not model any  electromagnetic counterparts. \cite{Petrov:2021bqm} repeats the analysis in \cite{Aasi:2013wya} with more realistic parameters and studies the resulting BNS lightcurves using only optical-wavelength kilonova models. Our study additionally takes into account uncertainty introduced by the use of realistic GW matched-filter pipelines, compares any potential differences between the use of low- and medium-latency skymaps, and is primarily focused on the infrared. We include only BNS mergers in this work and leave the study of NSBH mergers, which are expected to be especially promising to follow up in the infrared~\citep{Anand:2020eyg, Fernandez:2016sbf, Zhu:2020inc}, for future work. 

In Section \ref{sec:methods} we describe the simulation, including modeling a population of BNS mergers, their respective skymaps, and WINTER follow-up observations. We present the results in Section \ref{sec:results}. This leads to a discussion of the merits of studying kilonovae in the infrared and the planning of WINTER observations in Section \ref{sec:discussion}, and we conclude in Section \ref{sec:summary}.

\section{Methods} \label{sec:methods}

We simulate follow-up of BNS gravitational-wave signals with WINTER using the following procedure:
\begin{enumerate}
\item We generate a simulated BNS population and model a set of electromagnetic lightcurves from two different model grids~\citep{Kasen:2017sxr, Bulla:2019muo} for each merger event. 
\item To model the gravitational-wave response to each event, we simulate Advanced LIGO~\citep{TheLIGOScientific:2014jea}, and Virgo~\citep{TheVirgo:2014hva} detector noise and network configurations expected for O4. We also include the Japanese interferometer KAGRA~\citep{Somiya:2011np, Aso:2013eba, Akutsu:2017kpk}, which joined the global network of ground-based gravitational-wave detectors at the end of the third observing run.
\item We perform a low-latency search for gravitational-wave events in our simulated data using the \pycbc pipeline~\citep{Nitz:2018rgo, DalCanton:2020vpm}.
Low-latency skymaps are generated using the \bayestar software~\citep{Singer:2015ema} to obtain the localization uncertainty contour for each event recovered by \pycbc. Full source characterization is then performed using the \bilby parameter estimation pipeline~\citep{Ashton:2018jfp, Romero-Shaw:2020owr} to obtain medium-latency skymaps. Our simulated \pycbc results, \bayestar skymaps, and \bilby posteriors are publicly available on Zenodo at \cite{frostig_danielle_2021_5507322}.
\item We simulate WINTER follow-up observations searching the resultant \bayestar and \bilby skymaps. If WINTER successfully observes the true location of the event, we then verify that the event is bright enough to qualify as a statistically significant detection. 
\end{enumerate}

\subsection{Simulated BNS population} \label{subsec:bns}

The fourth observing run of the Advanced LIGO gravitational-wave interferometers~\citep{TheLIGOScientific:2014jea} is expected to have a duration of one year and include a four-detector network consisting of the two 
Advanced LIGO detectors~\citep{TheLIGOScientific:2014jea}, Advanced Virgo~\citep{TheVirgo:2014hva}, and KAGRA~\citep{Somiya:2011np, Aso:2013eba, Akutsu:2017kpk, Aasi:2013wya}. One major goal of this study is to determine how many BNS systems would be detected in gravitational waves and then successfully followed up with WINTER during O4. To this end, we simulate a realistic population of BNS systems that could be detected over the course of one year of observing with this detector network operating at its expected sensitivity during this observing run~\citep{noise_curves}. A (quasi-circular) BNS system is fully described by 17 binary parameters, $\boldsymbol{\theta}$, including the component masses, six-dimensional spins, and tidal deformabilities of the individual neutron stars, and the extrinsic parameters---the distance, inclination angle, right ascension, declination, polarization angle, and time and phase at coalescence.

We simulate systems following the mass distribution providing the best fit to the population of Galactic double neutron stars and the components of GW170817 as determined in \cite{Farrow:2019xnc}, which is parameterized in terms of the ``slow'' and ``recycled'' binary neutron star components. In the isolated channel for BNS formation (e.g., \citealp{Tauris:2017omb}), the recycled neutron star forms first and gets spun up to a period of $\sim 10-100~\mathrm{ms}$ via accretion from its companion~\citep{1982CSci...51.1096R, 1982Natur.300..728A, Heuvel:2017ziq}. Conversely, the second-born neutron star does not experience this period of accretion, and spins down to a ``slow'' period of $\mathcal{O}(1)~\mathrm{s}$ on a timescale of $\sim 1~\mathrm{Myr}$. The Galactic double pulsar PSR J0737-3039A/B serves as the canonical example for this type of system~\citep{Burgay:2003jj, Lyne:2004cj}.

\begin{figure}[ht!]
\centering
	\includegraphics[width=\linewidth]{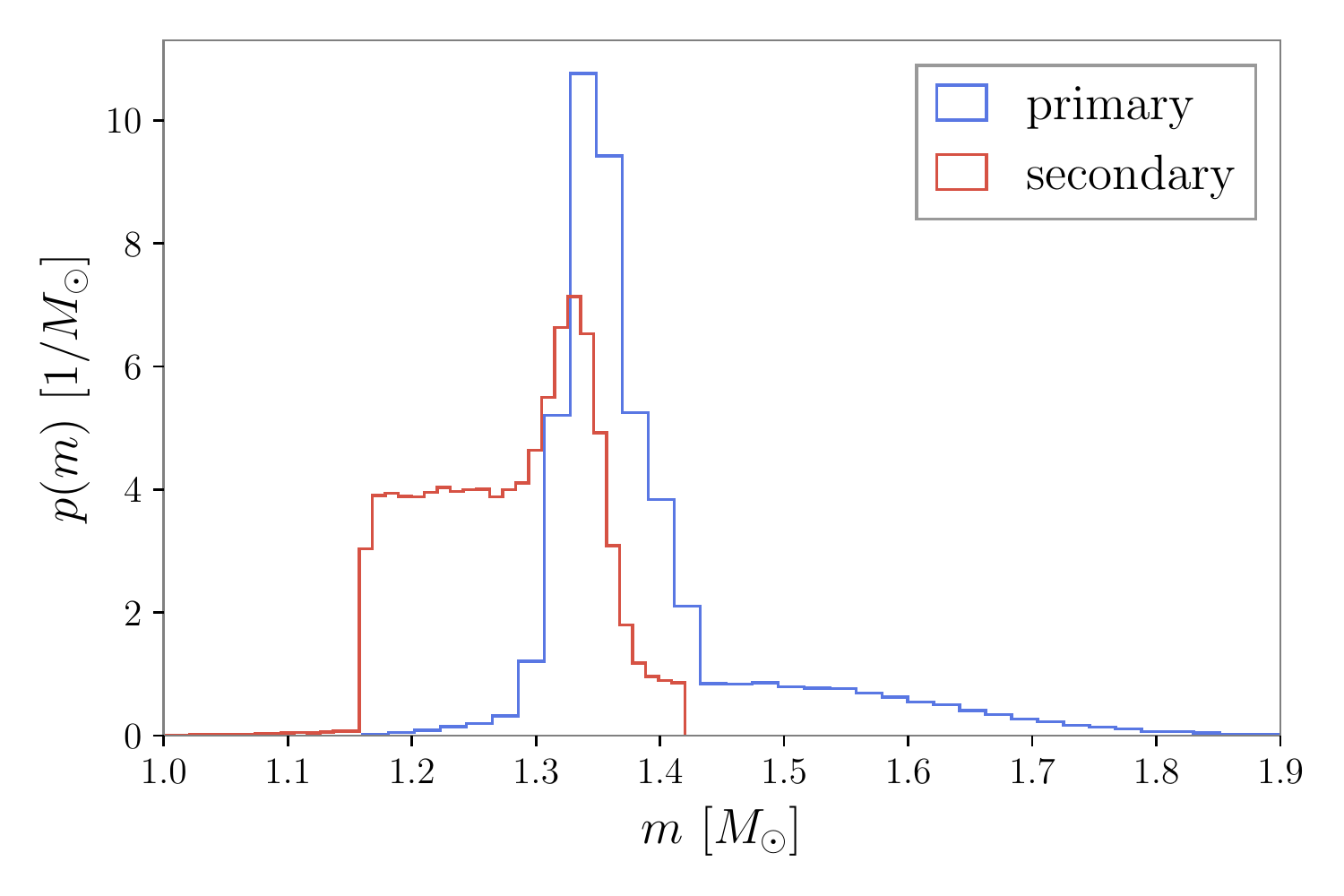}
	\includegraphics[width=\linewidth]{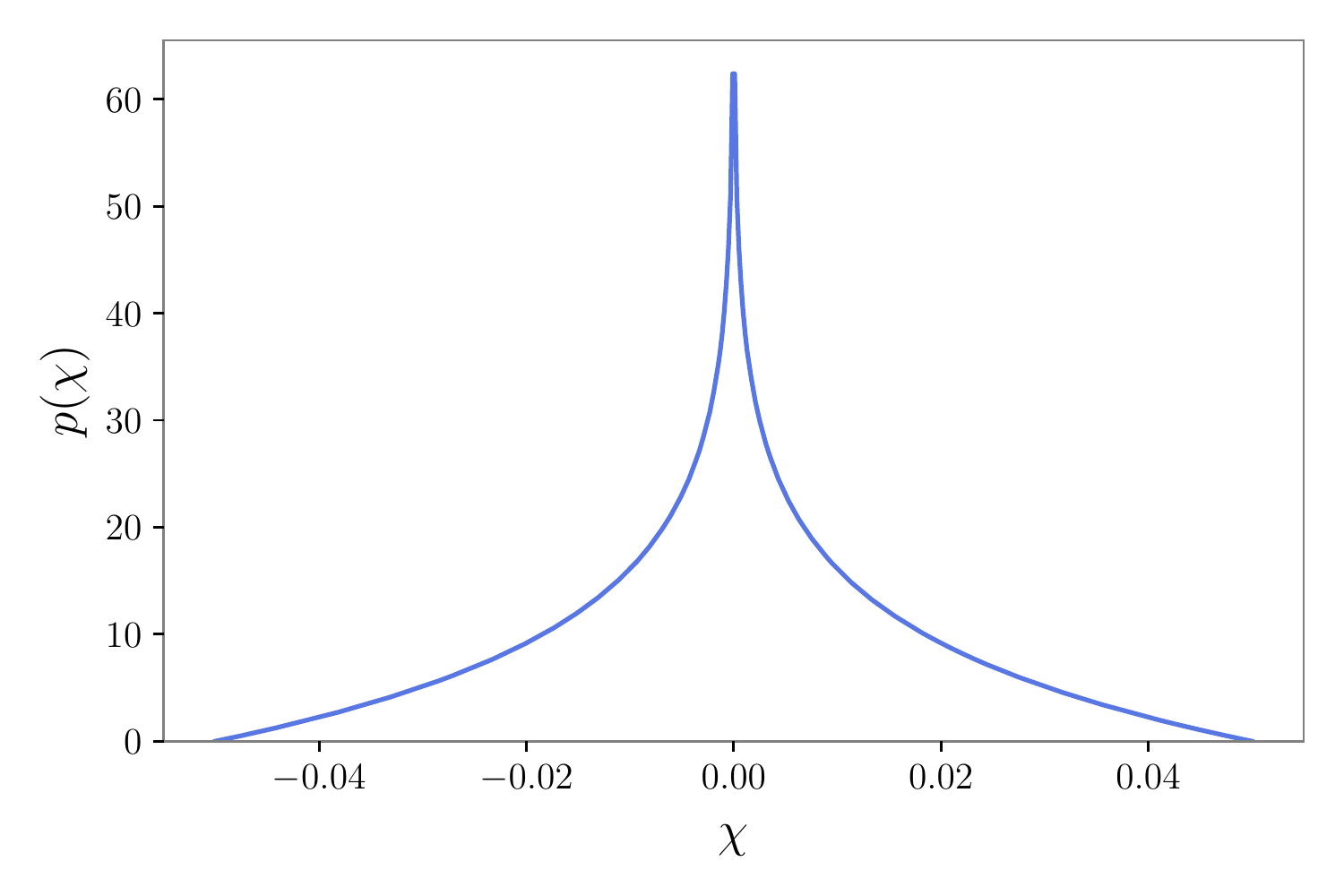}
\caption{\textit{Top}: Probability density function for the distribution of primary and secondary source-frame BNS masses from \cite{Farrow:2019xnc}. \textit{Bottom}: Probability density function for the distribution of dimensionless spins.}
\label{fig:intrinsic_distribution}
\end{figure}

We model the mass distribution of the recycled neutron star, $m_{r}$, as a Gaussian mixture model, with mixing fraction $\alpha=0.68$ determining the fraction of systems in the first Gaussian,
\begin{align}
    p(m_{s}) = \alpha\mathcal{N}(m_{r}; \mu_{1}, \sigma_{1}) + (1-\alpha)\mathcal{N}(m_{r}; \mu_{2}, \sigma_{2}).
\end{align}
The two Gaussians have means and widths given by $\mu_1=1.34~M_{\odot}, \sigma_1=0.02~M_{\odot}, \mu_2=1.47~M_{\odot},$ and $\sigma_2=0.15~M_{\odot}$.
We draw the masses of the slow neutron star, $m_{s}$, in each binary from a uniform distribution between $[1.16~M_{\odot}, 1.42~M_{\odot}]$~\citep{Farrow:2019xnc}. The primary and secondary masses are assigned via $m_1 = \max(m_r, m_s), m_2 = \min(m_r, m_s)$, the distributions for which are shown in the top panel of Figure~\ref{fig:intrinsic_distribution}. The tidal deformabilities of the neutron stars, $\Lambda_{i}$, are calculated from the masses assuming the AP4 equation of state~\citep{Akmal:1997ft, Akmal:1998cf}.

We assume the dimensionless spins of the neutron stars $\boldsymbol{\chi}$ are aligned with the orbital angular momentum and drawn from the implied distribution on $\chi_{z}$ assuming that the magnitudes are distributed uniformly on $[0, 0.05]$ and the directions are isotropic. This spin prior choice is motivated by the maximum spin of observed Galactic double neutron star systems~\citep{Lorimer:2008se}. The resultant distribution is shown in the bottom panel of Figure~\ref{fig:intrinsic_distribution}. The inclination angle, $\theta_{\mathrm{JN}}$, between the total angular momentum and the line of sight of the observer is drawn uniformly in $\cos\theta_{\mathrm{JN}}$, and the orbital phase at coalescence and polarization angle also follow uniform distributions.

The detector-frame coalescence times, $t_d$, of the sources are distributed uniformly over the course of one year starting on January 1\textsuperscript{st} 2022. We choose the sky locations and distances (and hence the redshifts) assuming the sources are distributed isotropically on the sky and uniformly in comoving volume and source-frame time, such that $p(z, t_d) \propto \frac{dV_{c}}{dz}(1+z)^{-1}$ out to a luminosity distance of $400~\mathrm{Mpc}$, where $V_{c}$ is the comoving volume. We consider three different merger rates consistent with those presented in \cite{LIGOScientific:2021psn}: pessimistic, realistic, and optimistic rates of $100,\ 500,$ and $1000~\mathrm{Gpc^{-3}{yr}^{-1}}$, respectively. This results in 21, 105, and 210 sources per year out to $d_{L}=400~\mathrm{Mpc}$ (assuming a merger rate unevolving with redshift; see Table~\ref{tab:chip}). 
We generate 625 unique BNS merger events following the distributions specified above. For each of the three rates considered, we draw 101 different realizations of the corresponding total yearly number of events from these 625 systems. This allows us to marginalize over many different realizations of the ``universe'' described by each merger rate and more realistically account for uncertainties.

\begin{table}
    \caption{Number of BNS mergers occurring and detected in gravitational-waves for each rate considered in this manuscript during one calendar year of O4. The first column indicates the total number of mergers, while the second column gives the median and 90\% symmetric credible interval on the number of systems detected in gravitational waves obtained by averaging over the 101 different realizations of BNS merger combinations. Also included are the number of unique pairs of events that are found in gravitational waves within one day and one week of each other.}
\label{tab:chip}
\begin{ruledtabular}
\begin{tabular}{l l l l l}
    Rate & Mergers & Found & 1 day & 1 week\\
    \midrule
    Pessimistic & 21 & $3^{+3}_{-2}$ & $0^{+0}_{-0}$ & $0^{+1}_{-0}$ \\
    Realistic & 105 & $16^{+6}_{-5}$ & $1^{+1}_{-1}$ & $6^{+7}_{-5}$ \\
    Optimistic & 210 & $33^{+7}_{-7}$ & $4^{+4}_{-3}$ & $26^{+19}_{-11}$ \\
\end{tabular}
\end{ruledtabular}
\end{table}

\subsection{Detection in gravitational waves}

\begin{figure}[ht!]
\centering
	\includegraphics[width=\linewidth]{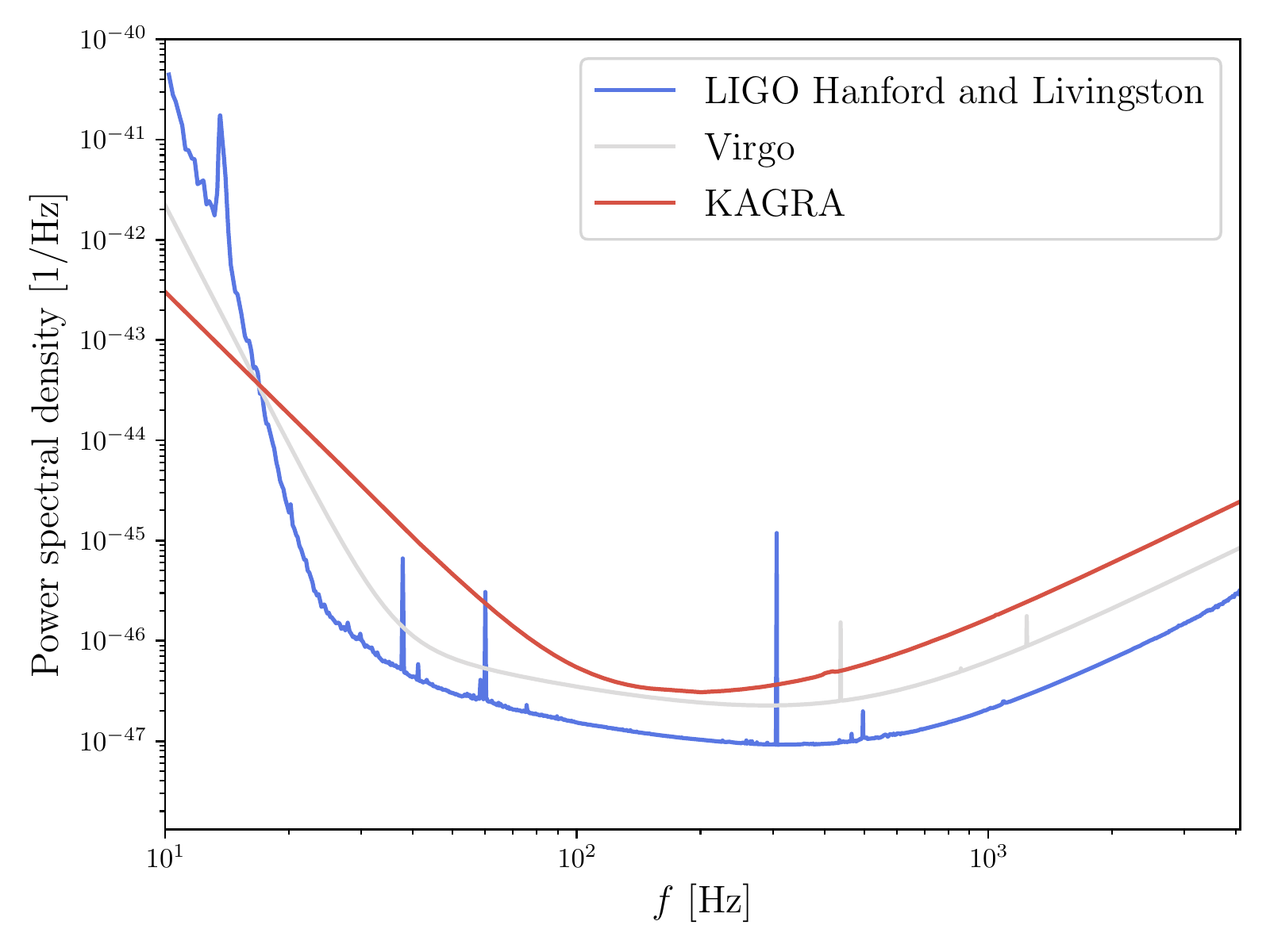}
\caption{Predicted power spectral densities during O4 of the four interferometers included in this study. These curves correspond to BNS ranges of 160-190 Mpc for the LIGO detectors, 90-120 Mpc for Virgo, and 80 Mpc for KAGRA~\citep{Abbott:2020qfu}.}
\label{fig:psds}
\end{figure}

For each simulated system, we randomly assign a detector configuration assuming each of the four interferometers are independently operating with a duty cycle of 0.7~\citep{Abbott:2020qfu}. Each BNS is added to $296~\mathrm{s}$ of simulated Gaussian noise sampled at $8192~\mathrm{Hz}$ colored by the expected O4 power spectral density (PSD, shown in Figure~\ref{fig:psds}) for each of the detectors that are ``observing'' during that event~\citep{noise_curves} starting at a frequency of $17~\mathrm{Hz}$.

\subsubsection{PyCBC Live}
\label{subsec:pycbc}

Low-latency searches for gravitational waves from compact binary mergers rely on the matched filtering technique (e.g. \citealp{Cutler:1992tc, Allen:2005fk}), where a template bank of waveforms is used to filter the data to search for candidate events. In order to more accurately account for the uncertainties associated with the matched-filter detection of gravitational-wave events, we use the \pycbc~\citep{Nitz:2018rgo, DalCanton:2020vpm} pipeline---which is one of the pipelines used to search for gravitational-wave events in LIGO and Virgo data in real time---to identify candidates in our simulated population. The discreteness of the template bank is the dominant source of uncertainty on the source parameter estimates obtained from matched filtering. It is constructed using a stochastic placement algorithm~\citep{Babak:2008rb, Harry:2009ea, Privitera:2013xza} so that the maximum loss in the signal-to-noise ratio ($\mathrm{SNR_{GW}}$) calculated from a starting frequency of 20 Hz between adjacent templates  is 3\%. The templates cover detector-frame component masses of between $1-3~M_{\odot}$ and aligned spins out to $\chi = 0.05$. 

\begin{figure}[ht!]
\centering
	\includegraphics[width=\linewidth]{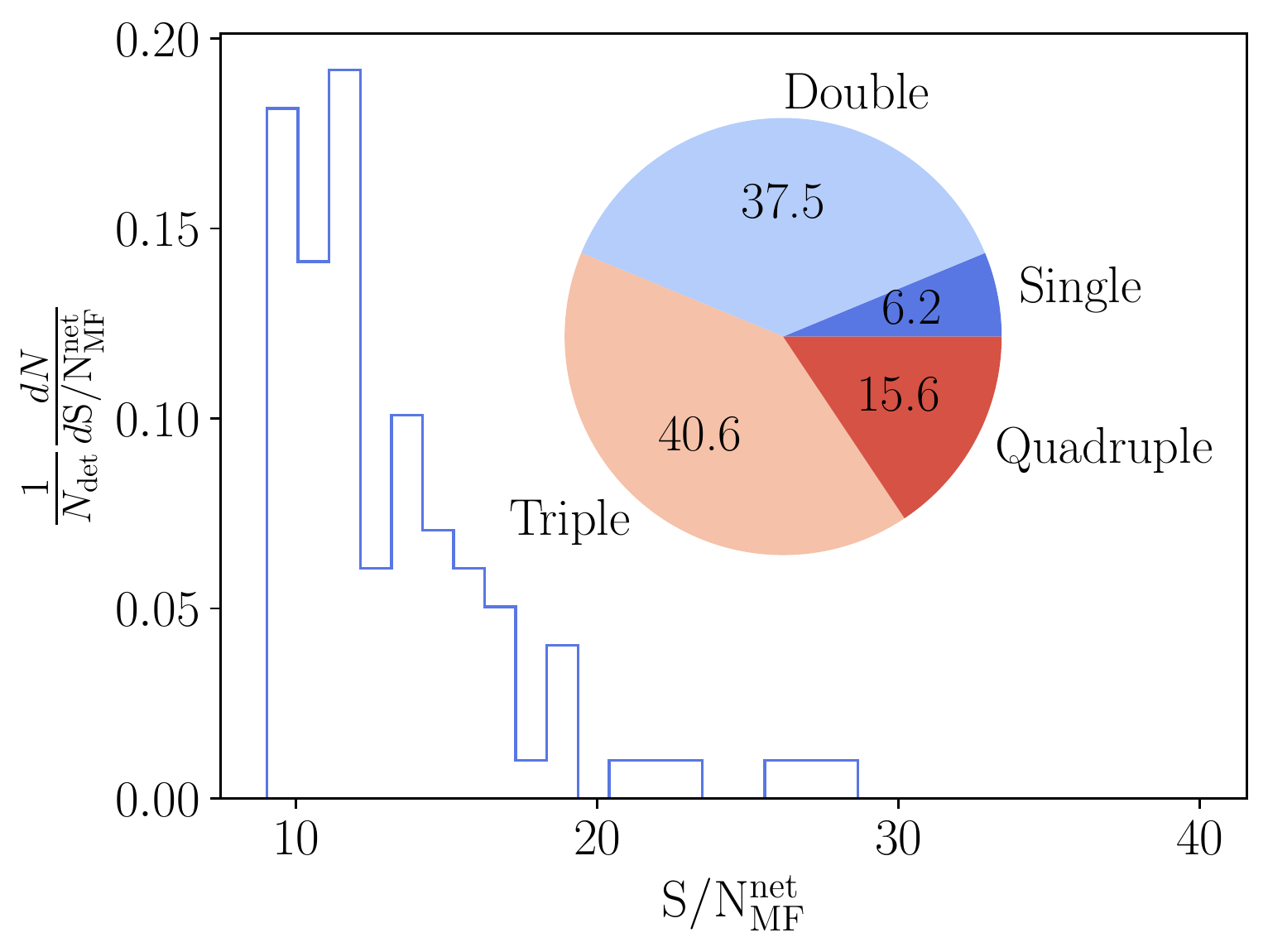}
\caption{Distribution of the network matched-filter $\mathrm{SNR_{GW}}$ recovered by \pycbc for the 96 total found events. The inset shows the fraction of the total number of found events detected with each number of interferometers.}
\label{fig:detector_networks}
\end{figure}

For convenience, the search is configured to run on data from all four detectors in our O4-like network. It calculates the PSD for each detector in real time using the simulated data for each event. A trigger is generated if the criterion $\mathrm{SNR_{GW}}\geq 4.5$ is met in at least two of the four detectors. In post-processing, we then calculate the network matched-filter $\mathrm{SNR_{GW}}$ using the trigger information returned by \pycbc by taking the square root of the quadrature sum of the $\mathrm{SNR_{GW}}$ in each detector that was actually ``observing'' at the time of the event in question.

Because we only simulate data in short segments around the times containing a BNS merger rather than a continuous one-year data stream, the false alarm rate (FAR) cannot be meaningfully calculated by \pycbc. Rather than using an FAR threshold to indicate detection, we instead set a threshold on the network matched-filter $\mathrm{SNR}_{\mathrm{GW}}$ recovered by the search for each event of $\mathrm{SNR}_{\mathrm{GW}} > 9$ in order for the source to count as ``found.''  Of the 625 independent BNS mergers, 96 are found by \pycbc. By averaging over the 101 different realizations of BNS merger combinations for each rate as described in the previous section, we obtain the median and 90\% symmetric credible interval on the number of found events shown in Table~\ref{tab:chip}. The fraction of the total number of events detected with each number of interferometers is shown in Figure~\ref{fig:detector_networks}, along with the recovered $\mathrm{SNR}_{\mathrm{GW}}$ distribution. The farthest event is found at a distance of $372.2~\mathrm{Mpc}$ and the nearest at $59.0~\mathrm{Mpc}$.

\subsubsection{\bayestar Map Generation} 
\label{subsec:bayestar}

We localize each event identified by \pycbc (i.e., with network $\mathrm{SNR}_{\mathrm{GW}} > 9$) using the \bayestar rapid localization algorithm \citep{Singer:2015ema}. During O3, skymaps produced using \bayestar were released in low-latency for each GW event that passed the public-alert threshold.

\bayestar takes as inputs the estimated masses of the neutron stars, the coalescence time, and the SNR time series from each detector as calculated by \pycbc. From this time series, \bayestar extracts the timing, relative phases, and amplitudes from each detector. \bayestar then constructs a three-dimensional sky localization posterior distribution, with a probability for each pixel on the two-dimensional sky and an additional distance component that is approximated by a Gaussian along each line of sight.

In O3 and in this work, \bayestar sky localizations for each event were computed in $\mathcal{O}$(10 s). Further efforts to optimize \bayestar have led to improvements resulting in runtimes of $\mathcal{O}$(1 s) \citep{Magee:2021xdx}.

\subsubsection{Parameter estimation}
After the initial low-latency skymap using the \bayestar algorithm is sent out, further source characterization is performed using full parameter estimation, whereby posterior probability distributions for the binary parameters are obtained using stochastic sampling methods. While \bayestar fixes the intrinsic parameter values to those corresponding to the maximum-likelihood template returned by the search pipeline, parameter estimation enables marginalization over the uncertainty in the intrinsic parameters and the extrinsic parameters that are not necessary for the skymap. Accounting for this uncertainty and allowing for correlations between parameters as done using full parameter estimation should lead to skymaps that have systematically smaller sky areas and include the true location of the source at smaller confidence intervals~\citep{Singer:2014qca}. We present a comparison of the skymaps from the two localization algorithms in Section~\ref{subsec:skymap_comp}.

The likelihood of observing gravitational-wave data $d$ given binary parameters $\boldsymbol{\theta}$ is given by~\citep{Veitch:2009hd, Romano:2016dpx, Ashton:2018jfp}:
\begin{align}
    \mathcal{L}(d | \boldsymbol{\theta}) \propto \exp\left( -\sum_{k} \frac{2|d_{k}-h_{k}(\boldsymbol{\theta})|^{2}}{TS_{k}}\right),
\end{align}
where $h(\boldsymbol{\theta})$ represents the gravitational waveform for the BNS signal with parameters $\boldsymbol{\theta}$, $T$ is the duration of the analyzed segment, $S_{k}$ is the PSD, and $k$ indicates the frequency dependence of the data, waveform, and PSD. The posterior probability distribution for the binary parameters characterizing each systems is then given by Bayes' Theorem:
\begin{align}
    p(\boldsymbol{\theta} | d) \propto \mathcal{L}(d | \boldsymbol{\theta}) \pi(\boldsymbol{\theta}),
\end{align}
where $\pi(\boldsymbol{\theta})$ represents the prior probability distribution assumed for $\boldsymbol{\theta}$.

We simulate the progression of skymaps that could occur during a real observing run by performing parameter estimation using the \bilby software~\citep{Ashton:2018jfp, Romero-Shaw:2020owr} for all of the events that were found by \pycbc. We use the \texttt{PyMultiNest} nested sampler~\citep{Feroz:2007kg, Feroz:2008xx, Feroz:2013hea, Buchner:2014nha} to generate samples from the posterior probability distribution for the source parameters, where the likelihood is analytically marginalized over the distance to the source and the phase at coalescence~\citep{Veitch:2014wba, Singer:2015ema, Thrane:2018qnx}. We make the assumption that the neutron stars are point masses with tidal deformability $\Lambda=0$, which is unlikely to affect the inference of the sky location of the source. This assumption enables the use of the reduced order quadrature implementation~\citep{Smith:2016qas} of the IMRPhenomPv2 waveform~\citep{Hannam:2013oca, Khan:2015jqa, Husa:2015iqa}, significantly reducing the computational cost of the parameter estimation. The likelihood uses the true PSD used to color the Gaussian noise in each detector. In order to keep the computational cost low, we do not marginalize over uncertainty in the detector calibration.

Because the search pipelines recover the chirp mass of the source extremely well~\citep{Biscoveanu:2019ugx}, we use a uniform prior on the detector-frame chirp mass with a width of $0.2~M_{\odot}$ centered on the true value. The prior on the mass ratio, $q\equiv m_{2}/m_{1}$, is uniform over the interval  $[0.125, 1]$. We restrict the analysis to aligned spins only, and the spin prior is the same as that used for drawing the simulated sources, shown in Figure~\ref{fig:intrinsic_distribution}. The luminosity distance prior is uniform in the source frame between 10 and 500 Mpc, and the prior on the coalescence time in the geocentric frame is uniform over a width of $0.2~\mathrm{s}$ centered on the true coalescence time. We use standard priors on all other parameters (see, e.g.,  \citealp{Romero-Shaw:2020owr}).

With these parameter estimation settings, the sampling stage takes $63^{+61}_{-32}~\mathrm{mins}$ across all of the recovered events. This depends most sensitively on the $\mathrm{SNR_{GW}}$ of the signal, since louder signals take longer to analyze due to the way the nested sampler explores the prior volume. We note that there are additional timing overheads for generating the weights for calculating the reduced order quadrature likelihood, reconstructing the posterior for the analytically marginalized distance parameter, and generating a skymap of the appropriate format that can be used by observers, although these additional steps generally take less time than the sampling. This timing is not necessarily indicative of what will occur in O4. There are a number of alternative techniques being explored that can be used to further accelerate parameter estimation, particularly for low-mass sources, like focused reduced order quadrature~\citep{Morisaki:2020oqk}, relative binning~\citep{Zackay:2018qdy, Finstad:2020sok}, parallelization~\citep{Pankow:2015cra, Talbot:2019okv, Wysocki:2019grj, Smith:2019ucc}, prior restrictions~\citep{You:2021eeq}, and machine learning based approaches~\citep{Gabbard:2019rde, Green:2020dnx, Green:2020hst, Williams:2021qyt}.

\subsection{Electromagnetic follow-up}
\subsubsection{Lightcurve modeling} \label{subsec:lightcurve}
In order to capture some of the current uncertainty in kilonova modeling, we use two different prescriptions to predict the lightcurves of kilonovae from our simulated BNS mergers.

\emph{Bulla model}: The first is a grid of kilonova models generated using the Monte Carlo radiative transfer code \texttt{POSSIS} \citep{Bulla:2019muo}. The grids assume a two-component ejecta comprised of mass ejected on a dynamical timescale and mass ejected in the form of a wind from the debris disk post merger. The dynamical ejecta is further divided into a lanthanide-rich equatorial component and a lanthanide-free polar component. The lightcurves are calculated using four parameters: the total dynamical ejecta mass ($M_{\rm{ej}}^{\rm{dyn}}$), the total wind ejecta mass ($M_{\rm{ej}}^{\rm{wind}}$), the opening angle of the lanthanide-rich component ($\Phi$) and the observer viewing angle ($\theta_{\rm{obs}} = \theta_{\rm{JN}}$ if $\theta_{\rm{JN}} < 90^{\rm{o}}$ and $\theta_{\rm{obs}} = 180^{\rm{o}} - \theta_{\rm{JN}}$ if $\theta_{\rm{JN}} > 90^{\rm{o}}$). A larger value of the opening angle $\Phi$ increases the amount of lanthanide-rich material, making the kilonova redder. Further details about the BNS grids are presented in \citet{Dietrich:2020efo}.

\emph{Kasen model}: The second model uses the kilonova simulations presented in \cite{Kasen:2017sxr}. While these simulations only include the effects of a single, spherically symmetric ejecta component with radial density given by a broken power law, they feature a more rigorous treatment of the microphysics determining the radiation transport in the system and allow for a range of compositions. The model parameters are the total ejecta mass, which we calculate as $M_{\rm{ej}} = M_{\rm{ej}}^{\rm{dyn}} + M_{\rm{ej}}^{\rm{wind}}$, the expansion velocity of the ejecta, $v_{\rm{ej}}$, and the mass fraction of lanthanides, $X_{\mathrm{lan}}$. A higher $M_{\rm{ej}}$ leads to lightcurves that peak at brighter magnitudes, while higher $v_{\rm{ej}}$ leads to faster-fading lightcurves. More neutron-rich ejecta with a higher lanthanide fraction, $X_{\mathrm{lan}}\gtrsim 10^{-2}$, produce a redder and longer-lived kilonova than ejecta composed primarily of light r-process material, $10^{-6} \lesssim X_{\mathrm{lan}} \lesssim 10^{-2}$~\citep{Kasen:2017sxr}. 

To estimate the ejecta masses and velocity from the neutron star masses, we use the prescription outlined in \citet{Stachie:2021noh}. First, we assume an AP4 equation of state~\citep{Akmal:1997ft, Akmal:1998cf} for the merging neutron stars to calculate their radii and compactness, although recent works have developed~\citep{Stachie:2021noh} and employed~\citep{LIGOScientific:2021qlt} methods for marginalizing over the uncertainty in the equation of state. We then calculate the dynamical ejecta mass using the fitting formula from \citet{Coughlin:2018fis}:
\begin{align}
    \log_{10}M_{\rm{ej}}^{\rm{dyn}} = \bigg[a\frac{(1 - 2C_1) ~m_1}{C_1} & + b ~m_2 \left(\frac{m_1}{m_2}\right)^{n} + \frac{d}{2}\bigg] \\ \nonumber
    & + [1 \leftrightarrow 2]
\end{align}
where $m_i$ and $C_i$ are the masses and the compactnesses of the two neutron stars, respectively, and $a = -0.0719$, $b = 0.2116$, $d = -2.42$ and $n = -2.905$. The ejecta velocity is similarly calculated using a fitting formula from \citet{Coughlin:2018fis}:
\begin{equation}
    v_{\rm{ej}} = \left[a^{\prime\prime}\frac{m_1}{m_2}(1+c^{\prime\prime}~C_1) + \frac{b^{\prime\prime}}{2}\right] + [1 \leftrightarrow 2],
\end{equation}
where the coefficients are given by $a^{\prime\prime}=-0.3090$, $b^{\prime\prime}=0.657$, and $c^{\prime\prime}=-1.879$.

 \begin{figure*}
    \centering
    \includegraphics[width=\textwidth, trim={1.5cm 0 1.5cm 0},clip]{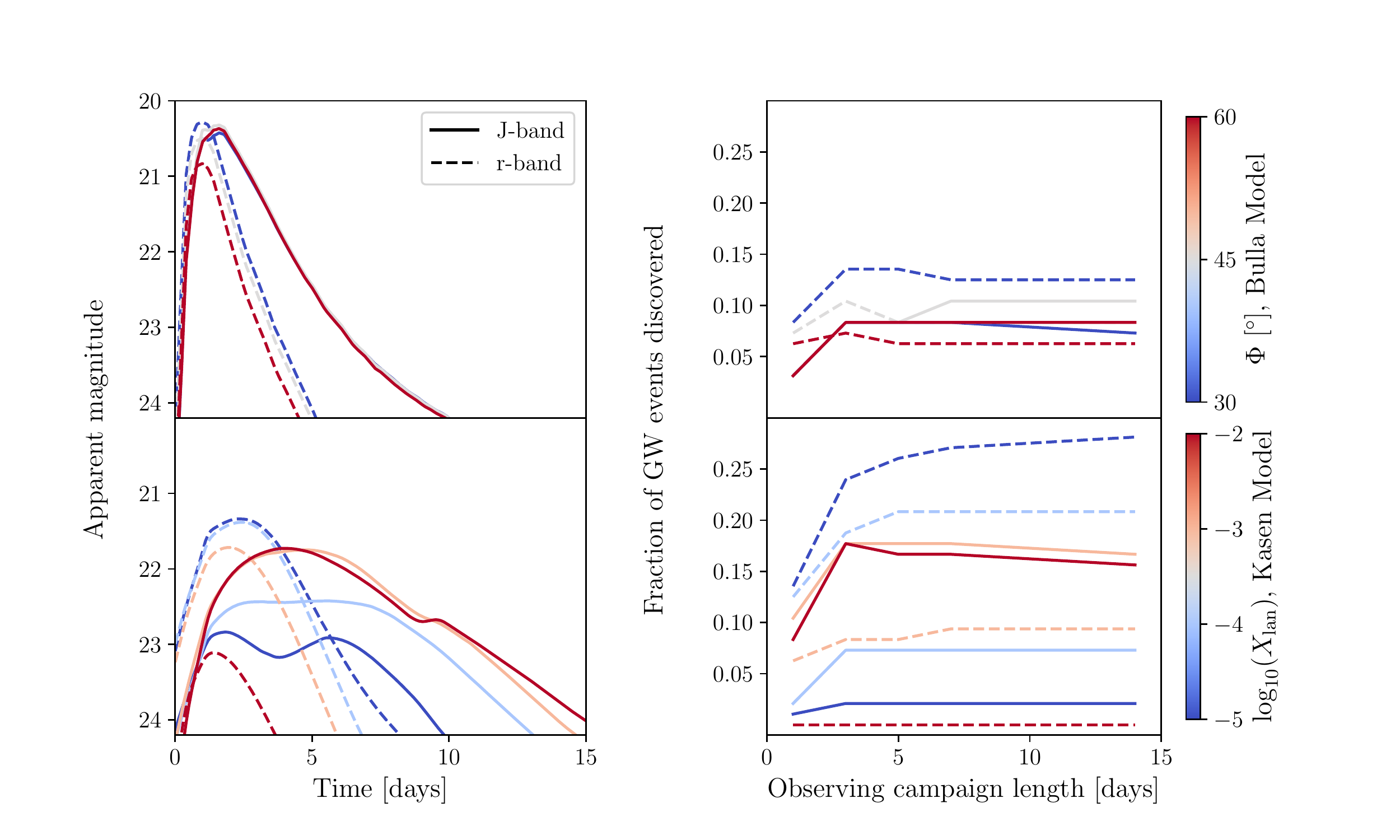}

    \caption{Variation due to lightcurve models. \textit{Left:} Example lightcurves in the J (solid lines) and r (dashed lines) filters for the Bulla model (top) with different opening angles and the Kasen model (bottom) with different lanthanide fractions as shown by the top and bottom color bars, respectively. The example BNS system has masses $m_1= 1.36~M_{\odot},\, m_2=1.22~M_{\odot}$ and is at a distance of $d_L= 162.97~\mathrm{Mpc}$.  \textit{Right:} The fraction of events discovered in the J band by WINTER and an equally sensitive r-band telescope, as described in Section~\ref{subsec:gwemopt}, plotted against the  number of nights allocated to a kilonova search, marginalized over all BNS merger event rates. Unique lightcurves are modeled for each event from the same set of Bulla (top) and Kasen (bottom) models.}
    \label{fig:Kasen_Bulla_lcs_discovery}
\end{figure*}

We estimate the wind ejecta mass as a fraction of the disk mass $M_{\rm{ej}}^{\rm{wind}} = \zeta M_{\rm{ej}}^{\rm{disk}}$ and set $\zeta = 0.15$ \citep{Dietrich:2020efo}. We calculate the disk mass using the fitting formula
\begin{align}
    &\log_{10}M_{\rm{ej}}^{\rm{disk}} =\\ \nonumber 
    &\max{\left[-3, a^{\prime} \left( 1 + b^\prime ~\tanh\left(\frac{c^{\prime} - (m_1 + m_2) / M_{\mathrm{thresh}} }{d^{\prime}}\right)\right)\right]}
\end{align}
where $M_{\rm{thresh}}$ is the minimum total mass that results in a prompt collapse post merger and is calculated as in \citet{Bauswein:2013jpa}; $a^{\prime}$, $b^{\prime} $, $c^{\prime}$, and $d^{\prime}$,  are calculated as in \citet{Dietrich:2020efo}.

We interpolate between the parameters of standard grids for both the Bulla and Kasen models using the Python package \gwemlightcurves \citep{Coughlin:2018miv, Coughlin:2018fis, Dietrich:2020efo} to predict the J-band and r-band lightcurves of the kilonovae. The left panel of Figure \ref{fig:Kasen_Bulla_lcs_discovery} shows the lightcurves for an example system. The differences in peak magnitude and decay time for the lightcurves calculated with the Kasen and Bulla models demonstrate the uncertainties currently underlying kilonova modeling. We present results for both models to conservatively account for this uncertainty in our predictions. Additionally, 
we note that the ejecta masses calculated from the fitting formulae are approximate and depend strongly on the equation of state, fitting parameters, and assumptions of the numerical relativistic models that they are based on. It is thus possible that the ejecta masses are not entirely representative of realistic kilonovae. We investigate the effect of varying ejecta masses in Section \ref{subsec:comparisons}.

\subsubsection{WINTER simulated observing} \label{subsec:gwemopt}

For each found gravitational-wave event, we use the corresponding skymap and simulated time of the event to create a realistic observing schedule with the \gwemopt package~\citep{Coughlin:2018lta, Coughlin:2019qkn, Ghosh:2015sxp}. The package takes gravitational-wave probability maps as inputs, such as the \bayestar and \bilby skymaps, subdivides the sky into tiles sized to the telescope field of view, and generates an optimized observing schedule. During O3, a network of telescopes including the Zwicky Transient Facility \citep{Bellm_2018} created GW follow-up schedules with {\tt\string gwemopt} (e.g.,  \citealp{Coughlin:2019qov}; \citealp{Kasliwal:2019pyx}) and we expect to use the package during nominal WINTER operations in O4.

We run a set of observing simulations allowing one, three, five, seven, and fourteen nights of dedicated telescope time searching for the kilonova from each event. For the scope of this study, we limit observations to the J-band to match WINTER's planned J-band reference images and all-sky survey and only study one gravitational-wave event at a time, even though multiple events may occur during the same night or same week (see Table~\ref{tab:chip}). Each WINTER observation lasts 450 seconds to match the $\text{J}_{AB}=21$ magnitude reference images, split across five dithers, with the time for dithering at approximately one second per dither represented as an overhead time in the simulation. Time to slew between each field is calculated based on the telescope and dome slew rates measured at WINTER's host telescope at Palomar Observatory. WINTER’s InGaAs sensors read out continuously during each exposure, leading to no overhead time due to sensor readout \citep{Malonis:2020}. Combining overhead and exposure times, for an eight hour night WINTER covers $\sim 63 ~\text{deg}^2$ to $\text{J}_{AB} = 21$ magnitudes. The WINTER data processing pipeline will subtract new science images from prebuilt reference images (constructed well before the GW alert) and detect candidate kilonovae in near real time.

We follow up each skymap with a ranked search strategy, in which we subdivide the sky into a fixed grid of telescope pointings. The center of each tile corresponds to a WINTER reference image for easy image subtraction during data processing. We prioritize the tiles based on the skymap-generated probability. Finally, we create a greedy observing schedule, where the highest probability tiles are observed first, as described in \cite{Coughlin:2018lta}. We prioritize scheduling at least two observations of each field, spread out over the length of an observing campaign to study the lightcurve evolving over time. In this simplified simulation, one observing schedule is created and executed without modification for each follow-up campaign. The simulation schedules multiple visits for each field, but fields are not preferentially revisited. During real O4 follow-up observing, WINTER will iteratively compare observations to prebuilt reference images and revisit any fields containing promising kilonova candidates (see Section~\ref{subsec:EM} for details)

Next, we check the observing schedules for a successful observation of the kilonova, which can be defined in multiple ways. For the sake of clarity in this study, we define an event as \textit{localized} if WINTER takes at least one image of the kilonova's true location on the sky. An event may not be localized if the skymap is too large for WINTER to search it efficiently or if the event is not overhead at the given time of year. Additionally, poor weather can hinder follow-up observations; however, weather simulations are outside of the scope of this study.

Even if an event is localized, the kilonova must be sufficiently bright at the time of imaging to be detected. The event qualifies as a \textit{discovery} if it is localized, imaged at least twice, and observed to a signal-to-noise ratio ($\text{SNR}_{\text{EM}}$) $\geq  5$. We employ the lightcurve models described in Section \ref{subsec:lightcurve} to calculate the magnitude of the event at the time of each observation and scale the magnitude based on the airmass of the observation. We follow Equation~1 as described in Section 3.1 of ~\cite{Frostig:2020} to calculate the $\text{SNR}_{\text{EM}}$ for each event based on historic Palomar data and predicted instrument noises. For comparison, we repeat this exercise for WINTER observing in J band and for a fictitious optical telescope observing in the Sloan Digital Sky Survey r filter with equivalent sensitivity, field of view, exposure times, and overhead times as the WINTER J-band observations.

\begin{figure}[ht!]
    \centering
    \includegraphics[width=0.45\textwidth]{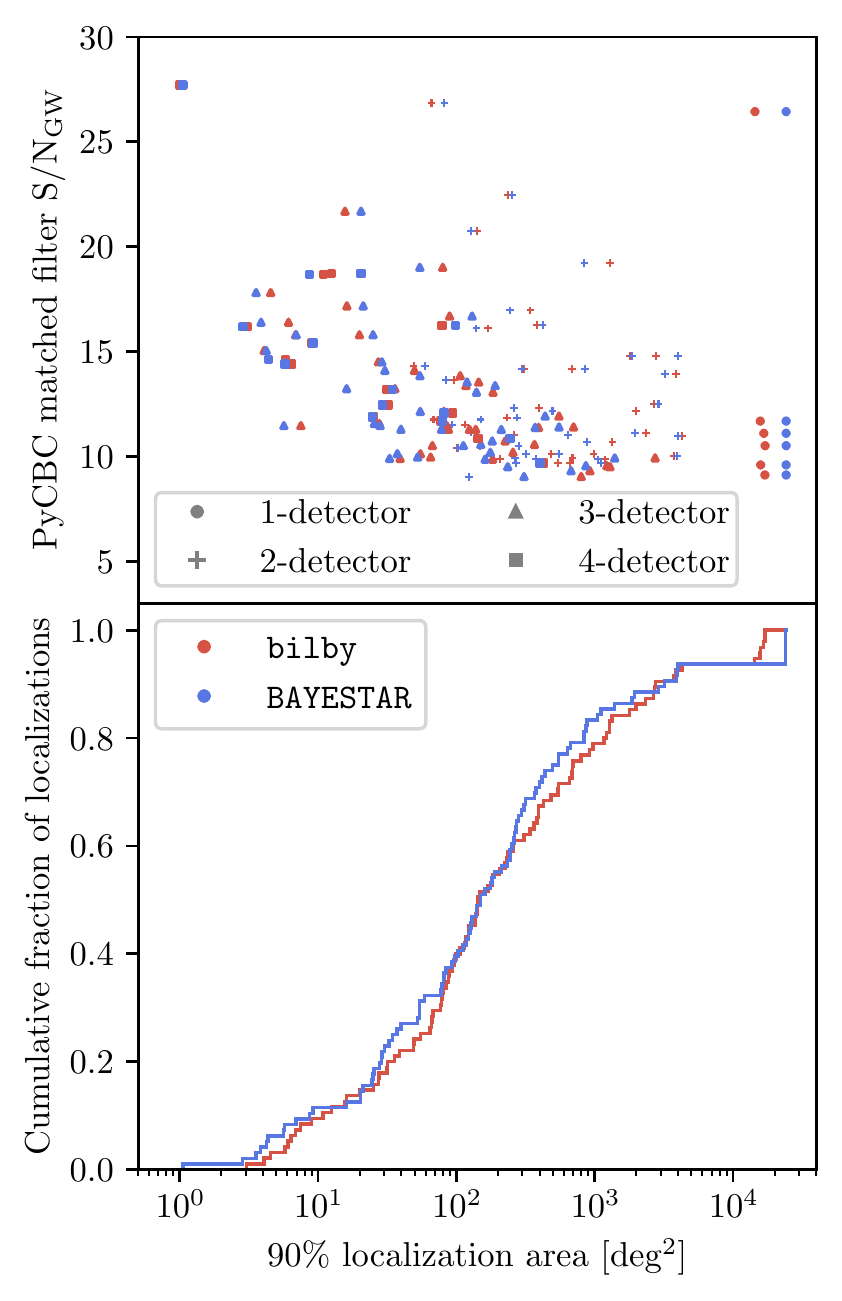}
    \caption{\textit{Top}: 90\% credible localization areas of \bilby and \bayestar skymaps for each event vs. its \pycbc matched filter $\text{SNR}_{\text{GW}}$. Different markers represent the number of GW detectors that identified the event.
    \textit{Bottom}: Cumulative distribution of 90\% credible localization areas for skymaps generated by both algorithms.}
    \label{fig:localization}
\end{figure}

For the scope of this study, we use two observations to $\text{SNR}_{\text{EM}}\geq 5$ as a simplified proxy for true kilonova discovery, as two observations are the minimum number required to distinguish a kilonova from asteroids or other transients. In reality, to confirm a new kilonova candidate, more than two observations may be required, and promising candidates will be selected based on their color evolution, how quickly the lightcurve fades, and if the event is associated with a probable host galaxy. Discovery will then be confirmed with further photometric and spectroscopic follow-up. See Section \ref{subsec:EM} for further discussion of  electromagnetic follow-up observations.

\section{Results} \label{sec:results}
\subsection{Skymap comparisons}
\label{subsec:skymap_comp}

\begin{figure*}
    \centering
    \includegraphics[width=\columnwidth]{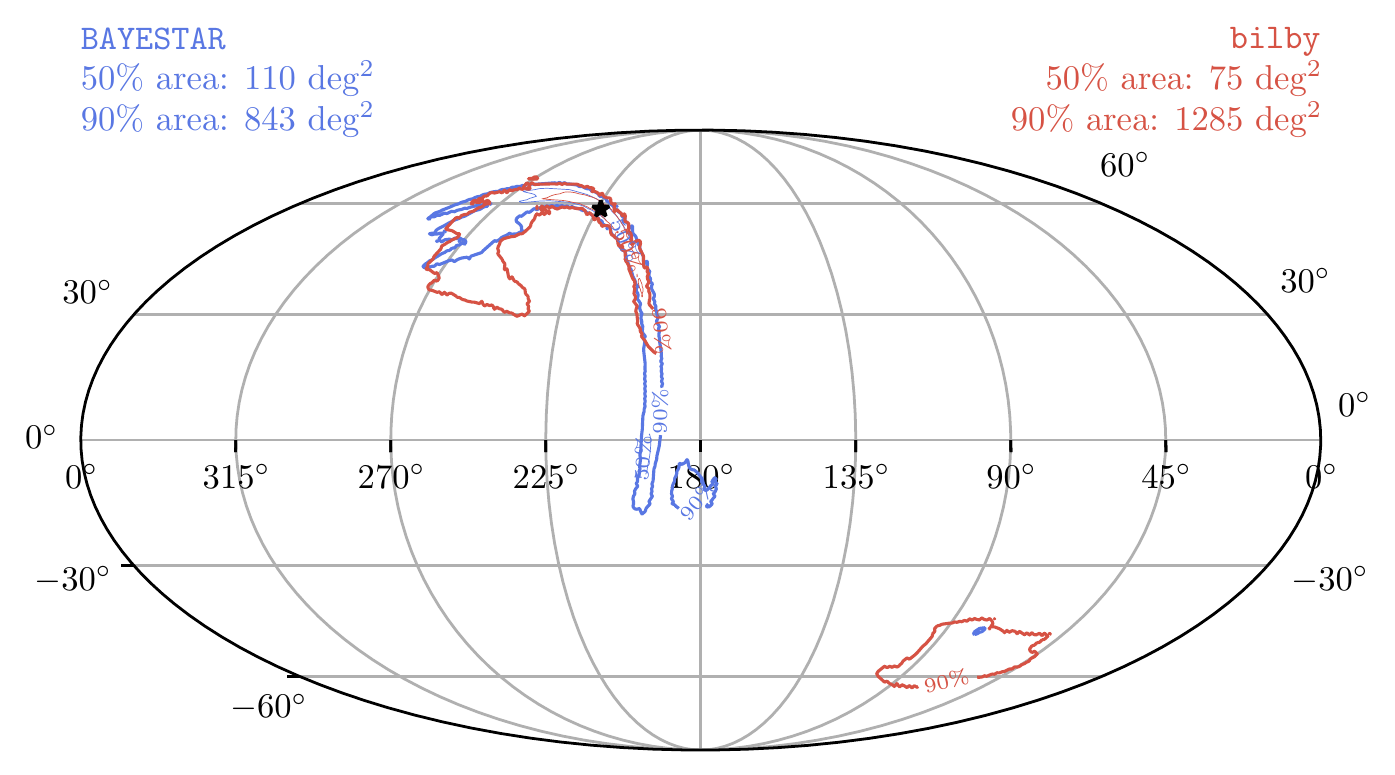}
    \includegraphics[width=\columnwidth]{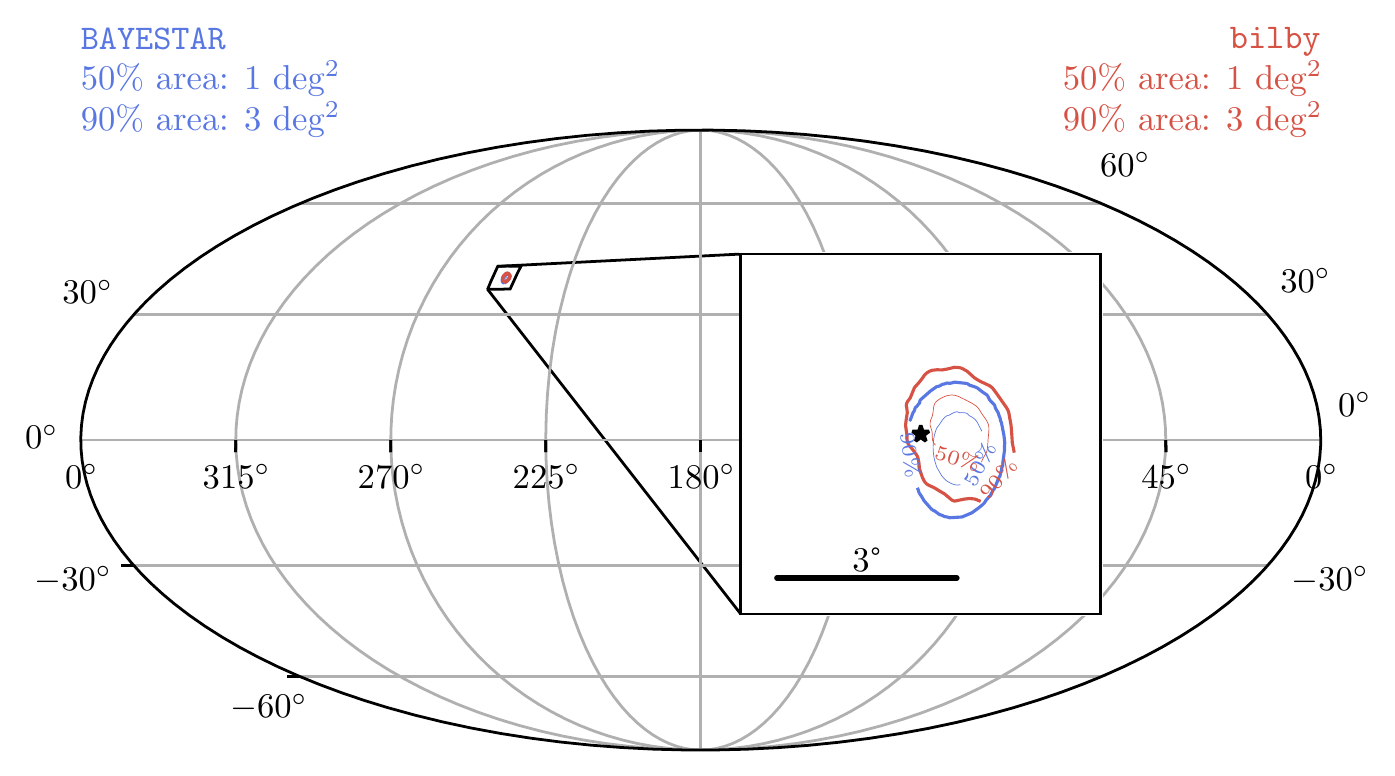}
    \caption{Comparison skymaps showing the 50 and 90\% credible regions obtained using \bilby and \bayestar for two different events. The true source location is marked with the black star. The event on the left was observed with the Hanford and KAGRA interferometers at a distance of 141 Mpc, although it was not confidently detected with KAGRA since the optimal SNR in that detector is 2.6. The one on the right was detected with Hanford, Livingston, Virgo, and KAGRA at a distance of 114 Mpc. In the skymap to the right, the \bayestar and \bilby localizations almost completely overlap.}
    \label{fig:skymap_comp}
\end{figure*}

\begin{figure*}
    \centering
    \includegraphics[width=\linewidth]{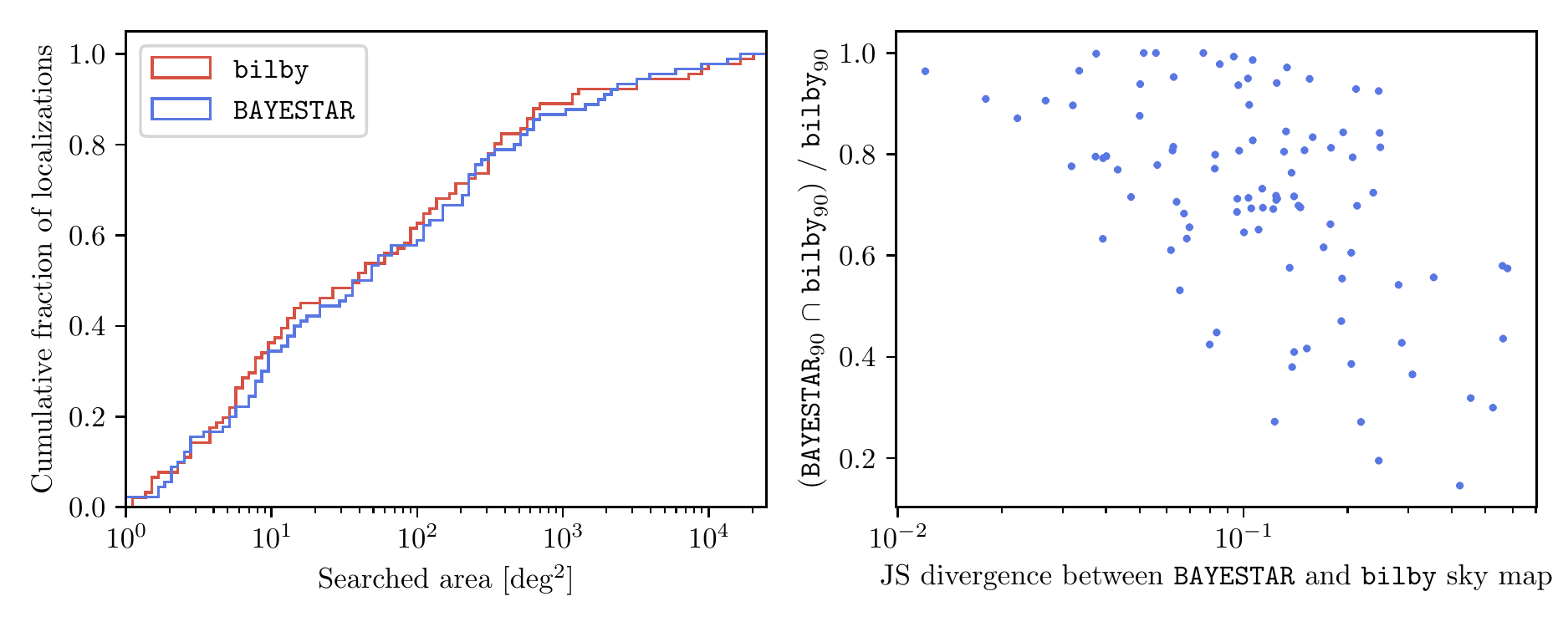}
    \caption{\textit{Left}: Cumulative distributions of searched area for skymaps generated using \bilby and \bayestar.
    \textit{Right}: The intersection of the \bayestar 90\% and \bilby 90\% credible areas, normalized by the \bilby 90\% credible area vs. Jensen-Shannon (JS) divergences between \bilby and \bayestar skymaps for each event}
    \label{fig:skymap_comp_plots}
\end{figure*}

A total of 96 events were found by \pycbc out of the 625 total independent simulated mergers, as shown in Table~\ref{tab:chip}. Each event was localized by both \bayestar and \bilby, as described in Section~\ref{sec:methods}. We present a comparison of the two algorithms. This is especially relevant for electromagnetic follow-up purposes, since in O3, \bayestar skymaps were  distributed hours or days before the more comprehensive \bilby or \lalinference~\citep{Veitch:2014wba} skymap. If this delay persists in O4 and if there are significant discrepancies in the two localizations, electromagnetic observers might switch partway through the night, or might even prefer to wait for the \bilby skymap before beginning observations.

The distribution of 90\% credible localization areas for all \bilby and \bayestar skymaps is shown in Figure~\ref{fig:localization}. Events that are recovered by \pycbc with larger values of matched-filter $\text{SNR}_{\text{GW}}$ tend to have smaller localization areas. Some events have network $\text{SNR}_{\text{GW}} > 20$ but localizations larger than 100~$\text{deg}^2$; these are typically one- or two-detector events where the low number of detectors results in poor constraints on timing and phase. The cumulative distributions of 90\% localization areas from each algorithm are similar, with 24\% of \bilby and 27\% of \bayestar localizations falling under 50 $\text{deg}^2$. The \bayestar distribution lies slightly to the left (i.e., to smaller areas) of the \bilby distribution.

Figure~\ref{fig:skymap_comp} shows a comparison between the \bayestar and \bilby skymaps for two particular events which are generally indicative of the performance observed in the larger sample. In the left skymap, which has a 90\% localization area of order $\mathcal{O}(1000) \text{deg}^2$ from a two-detector event at a distance of $d_{L}= 141.1~\mathrm{Mpc}$, the \bayestar and \bilby skymaps have some overlap, but with probability concentrated in different areas of the sky. In the right skymap, which stems from a four-detector event at a distance of $d_{L} = 113.8~\mathrm{Mpc}$, the localizations are small and in very good agreement.

A useful metric for the accuracy of localizations of simulated events is the searched area, which is the amount of sky area that is covered by integrating in order of decreasing probability from the highest probability pixel until reaching the true location of the source. The left panel of Figure~\ref{fig:skymap_comp_plots} shows the distributions of searched areas for \bilby and \bayestar. The two algorithms perform very similarly.

Another convenient statistic for comparing probability distributions is the Jensen-Shannon (JS) divergence. The JS divergence measures how similar two distributions are and ranges from 0 bit for identical distributions to 1 bit for completely divergent distributions. JS divergence values greater than 0.002 are considered to be statistically significant~\citep{Romero-Shaw:2020owr}. The right panel of Figure~\ref{fig:skymap_comp_plots} shows the intersection of the two 90\% credible areas normalized by the \bilby 90\% area versus the JS divergence between the two skymap posterior distributions for each event. These distributions describe the probability of the event being in each pixel of the two-dimensional skymap. For skymaps that have almost complete 90\% area overlap, the JS divergence is very small and thus the maps share significant information content, and vice versa. However, there are also events for which the normalized intersections are close to unity, but have relatively large JS divergences. This is due to two types of events: events that have similar 90\% localization areas, but with different probability distributions for the pixels within that area; and events for which the intersection of the \bilby 90\% area is small compared to the \bayestar 90\% area, so the intersection divided by the \bilby area is almost unity.

Previous studies have found that localizations from full parameter estimation pipelines such as \lalinference, when compared to those from \bayestar, have systematically smaller sky areas and include the true location of the source at smaller confidence intervals (see, e.g. Figure 3 of \cite{Singer:2014qca}). In this study we find that the discrepancy between the two algorithms is significantly reduced, since our \bayestar skymaps use data from all online detectors, while those in \cite{Singer:2014qca} use data only from detectors that register an $\mathrm{SNR_{GW}}$ above the detection threshold. Furthermore, since matched-filter searches recover the true chirp mass of BNS systems with extremely high accuracy, to within $\sim \mathcal{O}(10^{-4})~M_{\odot}$~\citep{Biscoveanu:2019ugx}, not marginalizing over the uncertainty in the mass parameters should lead to less significant biases in the low-latency skymap compared to higher-mass sources. We expect the difference between the skymaps obtained with the two algorithms to be more significant for NSBH sources, which will be explored in future work. 

\begin{figure}[!ht]
    \centering
    \includegraphics[width=0.48\textwidth]{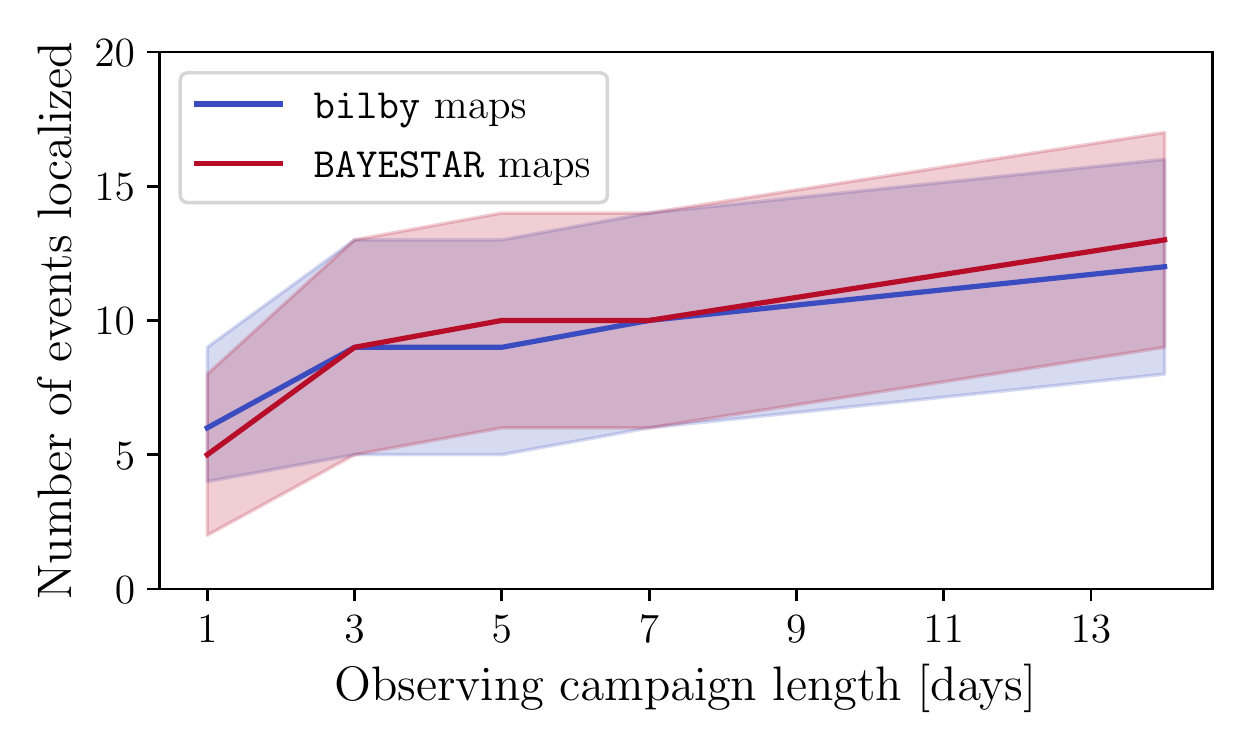}
    \caption{The number of events localized by WINTER for an optimistic BNS merger rate given the length of the observing campaign for following up each event with the \bilby and \bayestar skymaps. The solid lines represent the median number of events localized and the shaded regions show the 90\% symmetric credible interval.} 
    \label{fig:time_study}
\end{figure}

\subsection{Results from WINTER simulated observing}
\label{subsec:Winter_results}

Figure \ref{fig:time_study} shows the results of searching the \bilby and \bayestar maps with WINTER, varying the length of the observing campaign from one through fourteen nights of telescope time, given an optimistic BNS merger rate. The simulation produces separate observing schedules based on the amount of time allowed to follow up the event. In agreement with Section~\ref{subsec:skymap_comp}, WINTER localizes the BNS merger events at similar rates for both the \bilby and \bayestar skymaps. The number of events localized with WINTER steadily increases given more nights of observing time, with $\sim 2$ times more events localized with a fourteen-night search strategy than a one-night search strategy. 

\begin{table*}[!ht]
\centering
\caption{Number of gravitational-wave triggers leading to various categories of WINTER observations given pessimistic, realistic, and optimistic event rates. An event is accessible if it is overhead at Palomar Observatory at the given time of year, it is localized if the telescope takes an image of the kilonova's location on the sky, and the kilonova is discovered if detected at least twice to $\mathrm{SNR_{EM}}\geq 5$ in the J band. All simulations in this table use the ranked search follow-up strategy with five nights of searching the \bayestar skymaps.}
\label{tab:results}
\hspace*{-8.3em}
\resizebox{0.9\textwidth}{!}{%
\begin{tabular}{@{}lccccccc@{}}
\toprule[1pt]\midrule[0.3pt]

\textbf{Rate} &
  \multicolumn{1}{c}{\textbf{GW triggers}} &
  \multicolumn{1}{c}{\textbf{EM Accessible}} &
  \multicolumn{1}{c}{\textbf{Localized}} &
  \multicolumn{4}{c}{\textbf{Discovered}} \\
 &
  \multicolumn{1}{l}{} &
  \multicolumn{1}{l}{} &
  \multicolumn{1}{l}{} &
  \multicolumn{2}{c}{Bulla} &
  \multicolumn{2}{c}{Kasen} \\  \cmidrule(lr){5-6}  \cmidrule(lr){7-8}  
 &
  \multicolumn{1}{c}{Events} &
  \multicolumn{1}{c}{Events} &
  \multicolumn{1}{c}{Events} &

  \multicolumn{1}{c}{$\ \ \ \ \Phi\  [^{\circ}]$} &
  \multicolumn{1}{c}{Events} &
  \multicolumn{1}{c}{$\ \ \ \ X_{\mathrm{lan}}$} &
  \multicolumn{1}{c}{Events} \\ \midrule

Pessimistic &
  $3^{+3}_{-2}$ &
  $2^{+2}_{-2}$ &
  $1^{+1}_{-1}$ &

  \begin{tabular}[c]{@{}r@{}}30\\ 45\\ 60\end{tabular} &
  \begin{tabular}[c]{@{}r@{}}$0^{+1}_{-0}$\\ $0^{+1}_{-0}$\\ $0^{+1}_{-0}$\end{tabular} &
  \begin{tabular}[c]{@{}r@{}}$10^{-2}$\\ $10^{-3}$\\ $10^{-4}$\\ $10^{-5}$\end{tabular} &
  \begin{tabular}[c]{@{}r@{}}$0^{+2}_{-0}$\\ $0^{+2}_{-0}$\\ $0^{+1}_{-0}$\\ $0^{+0}_{-0}$\end{tabular} \\ \midrule
Realistic &
  $16^{+6}_{-5}$ &
  $11^{+5}_{-5}$ &
  $5^{+3}_{-3}$ &

  \begin{tabular}[c]{@{}r@{}}30\\ 45\\ 60\end{tabular} &
  \begin{tabular}[c]{@{}r@{}}$1^{+2}_{-1}$\\ $1^{+2}_{-1}$\\  $1^{+2}_{-1}$\end{tabular} & 
  \begin{tabular}[c]{@{}r@{}}$10^{-2}$\\ $10^{-3}$\\ $10^{-4}$\\ $10^{-5}$\end{tabular} &
  \begin{tabular}[c]{@{}r@{}}$2^{+3}_{-2}$\\ $3^{+2}_{-2}$\\ $1^{+2}_{-1}$\\ $0^{+1}_{-0}$\end{tabular} \\ \midrule
Optimistic &
  $33^{+7}_{-7}$ &
  $23^{+5}_{-7}$ &
  $10^{+4}_{-4}$ &

  \begin{tabular}[c]{@{}r@{}}30\\ 45\\ 60\end{tabular} &
  \begin{tabular}[c]{@{}r@{}}$3^{+1}_{-2}$\\ $3^{+2}_{-2}$\\ $3^{+2}_{-2}$\end{tabular} &
  \begin{tabular}[c]{@{}r@{}}$10^{-2}$\\ $10^{-3}$\\ $10^{-4}$\\ $10^{-5}$\end{tabular} &
  \begin{tabular}[c]{@{}r@{}}$6^{+3}_{-4}$\\ $6^{+4}_{-3}$\\ $2^{+2}_{-2}$\\ $1^{+1}_{-1}$\end{tabular} \\ 
 \midrule[0.3pt]\bottomrule[1pt]
\end{tabular}%
}
\end{table*}

Varying the number of nights allowed searching for the kilonova changes the ordering and prioritization of each observing schedule. For example, given a one-night observing campaign, 100\% of the images of the kilonova are taken before the J-band peak of the lightcurve, regardless of model. For a five-night campaign, 80\% (86\%) of the localized events have their first images taken before the peak and 30\% (27\%) have two images taken before the peak for the Kasen (Bulla) lightcurve model. Increasing observing time allows WINTER to cover a greater portion of the skymap, but risks tiling the high-probability area of the skymap less efficiently or observing the kilonova later in its evolution when it may have already faded. Therefore, discovery of new kilonovae (defined above as observing the event at least twice to $\text{SNR}_{\text{EM}}\geq5$) does not increase steadily with time, but levels off as the kilonova eventually fades beyond detection. In the right panel of Figure \ref{fig:Kasen_Bulla_lcs_discovery}, we compare discovery of new kilonovae in the WINTER J band and for an equally sensitive r-band telescope given the number of nights each telescope is allowed to search for the kilonovae. 

Given the Kasen lightcurve models, discovery in both J band and r band peaks around three nights of tiling and levels off given more nights of observing, with the exception of lanthanide-poor ($X_{\mathrm{lan}}= 10^{-5}$) kilonova discovery in the r band continuing to increase with fourteen nights of observing (bottom right panel of Figure \ref{fig:Kasen_Bulla_lcs_discovery}). In the J band with three nights of observing, WINTER discovers 8.5 times more lanthanide-rich ($X_{\mathrm{lan}}= 10^{-2}$) events than lanthanide-poor events ($X_{\mathrm{lan}}= 10^{-5}$). In contrast, an equally sensitive r band telescope discovers zero lanthanide-rich events and up to 28\% of lanthanide-poor events.

Despite differences in the lightcurve models, J-band and r-band discovery also levels off after three nights of observing for the Bulla models (top right panel of Figure \ref{fig:Kasen_Bulla_lcs_discovery}). Varying the opening angle of the lanthanide-rich component changes the resultant lightcurves less than varying the lanthanide fraction in the Kasen models. At three nights of observing, a large opening angle of the lanthanide-rich component ($\Phi=60^{\circ}$) decreases r-band discovery by 46\% and does not change J-band discovery.

Table \ref{tab:results} displays the results of the end-to-end simulation studying how many kilonovae WINTER will discover during one year of follow-up observations, including the median and 90\% symmetric credible interval on the number of events as described in Section~\ref{subsec:pycbc}. Some subset of the GW triggers will not be observable by a telescope at Palomar Observatory at the given trigger time, either because the event is too far south or too near the sun. For a given year of GW triggers, on average $\sim70$\% of the events are visible above $20^{\circ}$ altitude for WINTER at some point during the night following the event trigger time. These events are denoted as EM accessible in the table. For a portion of those events, WINTER observes the correct patch of sky and localizes the events, with some events missed due to large localization areas or events with true locations outside of the 90\% localization areas. Localized events only qualify as a discovery ($\mathrm{SNR_{EM}}\geq 5$ in at least two observations) if the event is sufficiently bright, which can vary based on the model grid used to simulate the event. WINTER discoveries range from as low as zero new kilonovae per year with a pessimistic BNS merger rate to as high as ten new kilonovae discovered per year to 90\% confidence with an optimistic BNS merger rate. Given a realistic BNS merger rate, WINTER discovers up to five new kilonovae per year to 90\% confidence.

\section{Discussion} \label{sec:discussion}
\subsection{Advantages of infrared follow-up} 
\label{subsec:comparisons}

\begin{figure*}
    \centering
    \includegraphics[width=\textwidth]{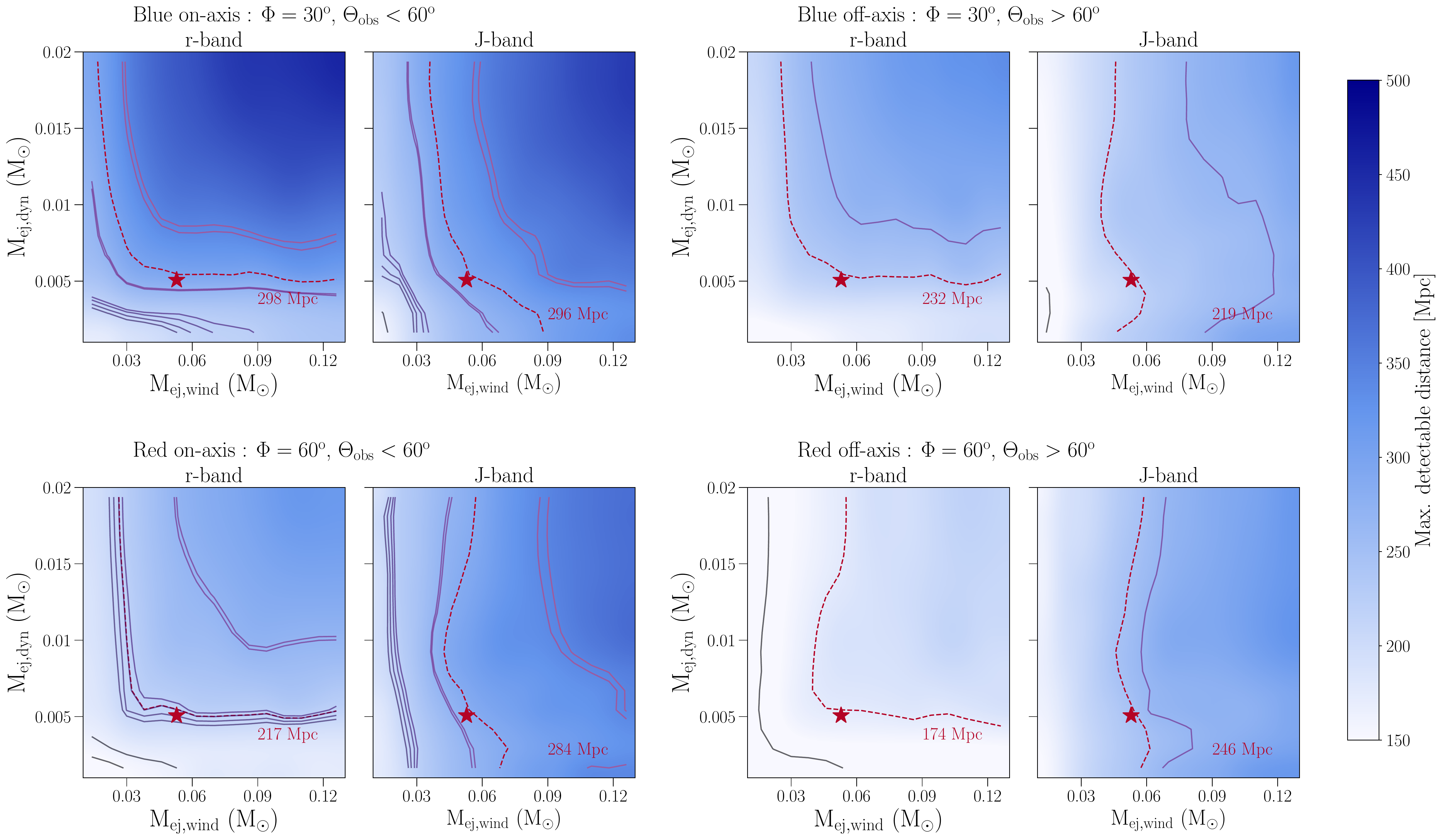}
    \caption{The maximum distance out to which a kilonova can be detected by an $m_{\rm{lim}}=21$ survey for a given combination of dynamical and wind ejecta masses, based on the Bulla models. We plot the distances separately for J and r bands. We further distinguish the kilonova models as blue and on-axis (top left), blue and off-axis (top right), red and on-axis (bottom left) and red and off-axis (bottom right). The ejecta masses for GW170817 are plotted as a red star. The red dashed line shows the contour at the distance at which a kilonova with GW170817-like ejecta masses is detectable in each case (distance indicated in red). We also plot the contours at the distances of 15 events followed up by WINTER from one realization of our realistic-rates simulation. Twelve of these events are on-axis and three are off-axis. Two off-axis events have distances $<150$ Mpc and lie off the plots. It is evident that for the same set of ejecta masses, the J band can detect kilonovae out to larger distances than r band if the kilonova is red.}
    \label{fig:ejecta_masses_lcs_comparison}
\end{figure*}

In Section \ref{sec:results}, we demonstrate that a 1 $\text{deg}^2$ J-band survey like WINTER can detect up to ten kilonovae during O4. Here, we examine the advantages of a J-band search over a similar optical search. We use a grid of realistic kilonova lightcurves calculated with the Bulla model to identify the parameter space where an infrared search outperforms an optical search. To generate the grid, we do not use the fitting formulae from Section \ref{subsec:lightcurve}, as they may not be entirely representative of the underlying physical population of kilonovae. Instead, we calculate lightcurves for a wide range of possible ejecta masses derived from numerical relativistic simulations: $M_{\rm{ej}}^{\rm{dyn}} = [0.001, 0.01]~M_{\odot}$ and $M_{\rm{ej}}^{\rm{wind}} = [0.01, 0.13]~M_{\odot}$ \citep{Andreoni:2020ewy}. We set the dynamical ejecta opening angles in the range $\Phi = [30^{\circ}, 60^{\circ}]$ and calculate lightcurves for viewing angles sampled uniformly in cos($\theta_{\rm{obs}}$).

\subsubsection{Red kilonovae are brighter at infrared wavelengths}
Kilonovae with a larger opening angle of the dynamical ejecta will have more lanthanide-rich material and hence will be brighter at redder wavelengths. We quantify this in Figure  \ref{fig:ejecta_masses_lcs_comparison} which shows the maximum distance out to which a kilonova can be detected in the r and J bands by a telescope with a limiting depth of 21 magnitudes. We distinguish between blue ($\Phi = 30^{\circ}$) and red ($\Phi = 60^{\circ}$) kilonovae, and an on-axis ($\theta_{\rm{obs}}< 60^{\circ}$) and off-axis ($\theta_{\rm{obs}}> 60^{\circ}$) kilonova. We use the maximum-likelihood estimates of the ejecta masses of GW170817 as a benchmark ($M_{\rm{ej}}^{\rm{dyn}} \sim$ 0.005 $M_{\odot}$,  $M_{\rm{ej}}^{\rm{wind}} \sim$ 0.05 $M_{\odot}$, \citealp{Dietrich:2020efo}) and indicate the distances out to which a kilonova with these masses can be detected. Finally, we plot contours corresponding to the distances of 15 events from one realization of our uniformly distributed, realistic-rate simulation that were detected in GWs and followed up with WINTER (see Table \ref{tab:chip}). Twelve of these events are on-axis and three are off-axis.

\begin{figure*}
    \centering
    \includegraphics[width=\textwidth]{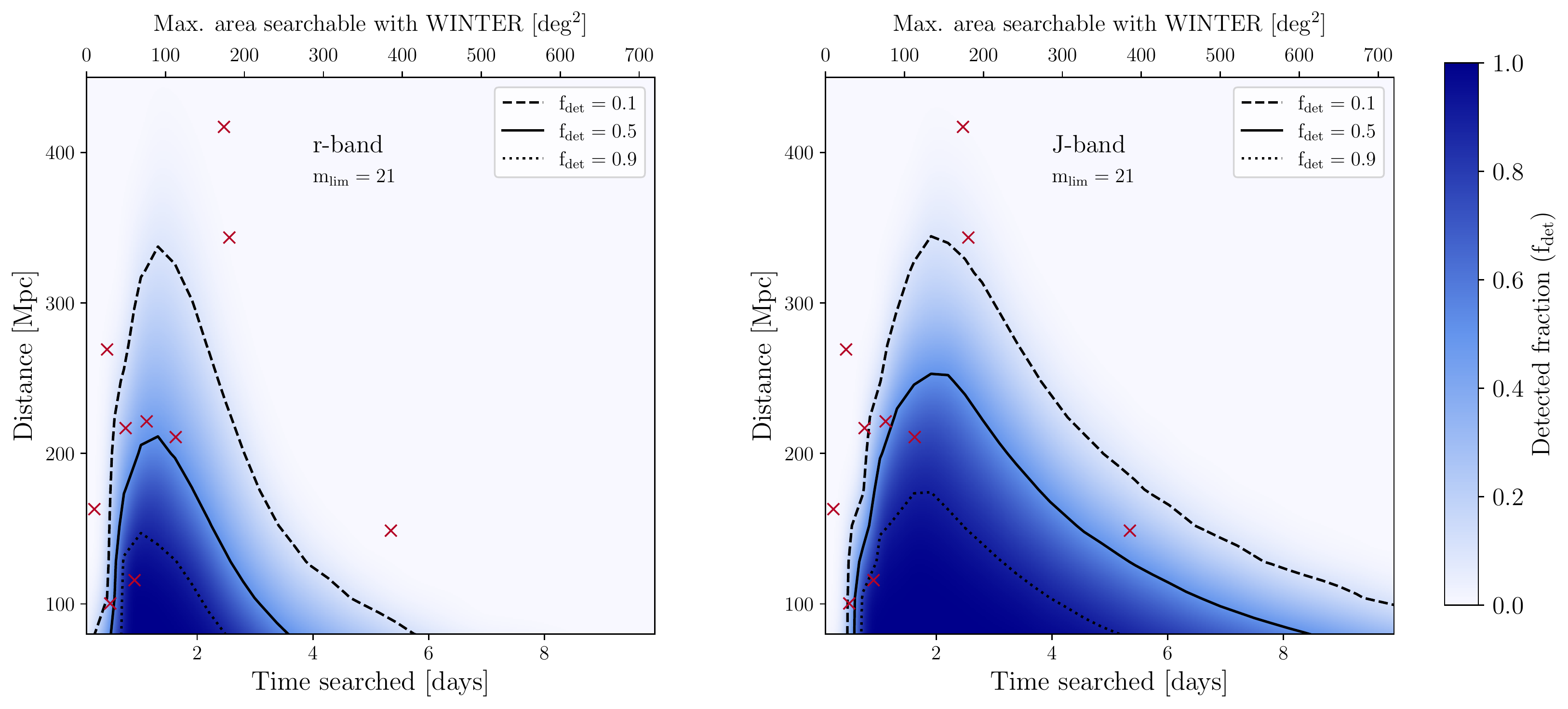}
    \caption{The fraction of kilonovae from our entire model grid that can be detected in the r and J bands by an m$_{\rm{lim}} = 21$ telescope, as a function of distance and number of days since the merger. The contours corresponding to detection fractions (f$_{\rm{det}}$) of 0.1, 0.5, and 0.9 are plotted as dashed, solid, and dotted lines, respectively. Red crosses mark the distances and areas enclosing 90\% localization probability of the events from our realistic-rate simulation (of the 16 events, only 10 are shown, as the remaining six lie outside the bounds of the axes). It is clear that kilonovae can be detected for much longer in the J band compared to r band. The right panel can also be used to select GW triggers that are worth following with WINTER. We will only follow events that have median distance estimates and localization areas such that the detection fraction is at least 10\%. We note that the detection fraction drops at very early search times ($t<0.5$ day), as the kilonova is still brightening in our models at these times. However, we will observe events with localization areas that can be tiled within 0.5 day, observing them repeatedly until the kilonova becomes bright enough to be detected.}
    \label{fig:time_distance_fraction}
\end{figure*}

For a blue kilonova ($\Phi = 30^{\circ}$), an infrared search does not offer a significant advantage over an optical search. A blue on-axis (off-axis) kilonova with GW170817-like ejecta would be detectable out to $\approx 300$ (230) Mpc in both r and J bands.  This is because the peak values of r- and J-band lightcurves for a blue kilonova are not significantly different (see top left panel of Figure \ref{fig:Kasen_Bulla_lcs_discovery}). Off-axis kilonovae are generally fainter than on-axis ones, which explains the reduced sensitivity (300 Mpc vs. 220 Mpc) for off-axis kilonovae. All 12 on-axis events from our simulation are detectable in both r and J bands if they are blue. We note that the detectable region of the ejecta mass phase space (i.e., the region to the right of each contour in Figure \ref{fig:ejecta_masses_lcs_comparison}) is slightly larger for r band than for the J band. Ten of these events have a detectable region that includes GW170817 ejecta masses in both r and J bands. All three off-axis events from our simulation are detectable in both r and J bands. Of these three, two are closer than 150 Mpc, and are detectable for the entire range of ejecta masses in both r and J bands.

However, an infrared search performs significantly better than an optical search for red kilonovae. Both on- and off-axis red kilonovae are detectable out to larger distances in the infrared than in the optical. A red on-axis (off-axis) kilonova with GW170817-like ejecta is detectable to 284 (246) Mpc in the infrared but only to 217 (174) Mpc in the optical. If the 12 on-axis kilonovae from our simulations are red, only 10 are detectable in the r band while all 12 are detectable in the J band. GW170817 ejecta masses lie in the detectable region for only six kilonovae in the r-band but for 10 kilonovae in J band. If the three off-axis simulated events are red, only two are detectable in the r band while all three can be detected in the J band.  
Finally, we note that if a particular kilonova is not detected in WINTER observations, Figure \ref{fig:ejecta_masses_lcs_comparison} can be used to place constraints on the ejecta masses and opening angles associated with it.

\subsubsection{All kilonovae are longer-lived in the infrared}

A second advantage of infrared searches is that kilonova emission is longer-lived in the infrared than in the optical. We quantify this in Figure \ref{fig:time_distance_fraction}, which shows the fraction of kilonovae from our entire model grid that can be detected in the r and J bands by an m$_{\rm{lim}} = 21$ telescope as a function of distance and number of days since the merger. We plot contours corresponding to detection fractions (f$_{\rm{det}}$) of 0.1, 0.5, and 0.9.

\begin{figure*}
    \centering
    \includegraphics[width=0.5\textwidth]{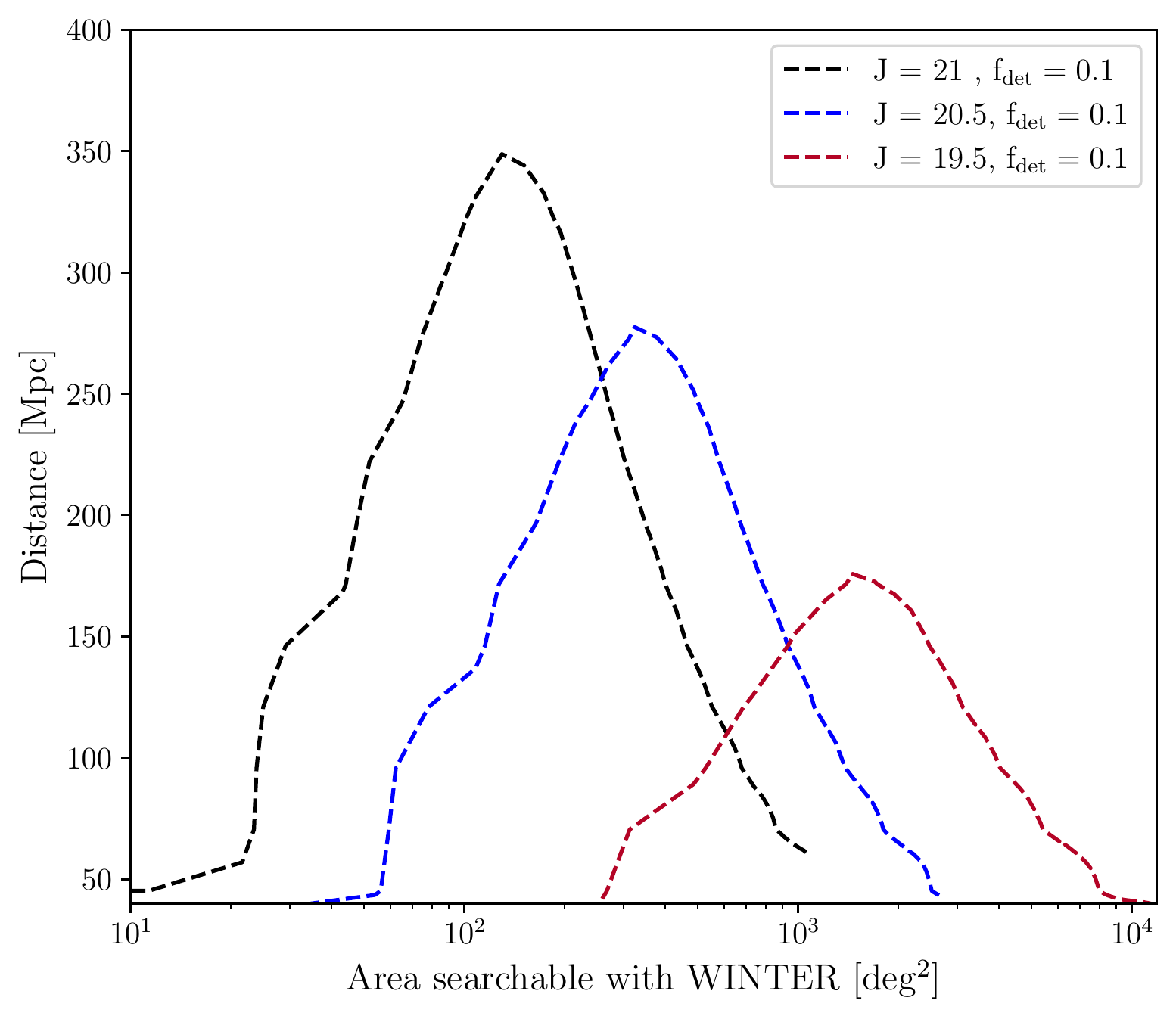}
    \caption{Contours marking the regions for which the kilonova detection fraction is at least 0.1 (f$_{\rm{det}}>0.1$), as a function of the localization area. We plot separate contours for searches with limiting magnitude of 21, 20.5, and 19.5 mag (black, blue, and red, respectively). With WINTER, we will follow up all events with localization areas $<450 $ $\text{deg}^2$. down to a depth of 21 mag. If the localization area is larger than 450 $\text{deg}^2$, we will follow only those events that have median distance estimates $< 200$ Mpc down to a depth of 20.5 mag. If the localization area is larger than 1000 $\text{deg}^2$, we will follow only those events with median distances $<150$ Mpc down to a depth of 19.5 mag.}
    \label{fig:time_distances_depths}
\end{figure*}

Figure \ref{fig:time_distance_fraction} clearly shows that kilonovae can be detected for much longer in the J band compared to r band. For example, at 200 Mpc more than 10\% (i.e. f$_{\rm{det}}>0.1$) of kilonovae are detectable in the r band for a maximum duration of three days. However, in the J band the same events can be detected for almost six days. A WINTER-like telescope with a 1 $\text{deg}^2$ field of view can thus search localization regions of $\approx 450$ $\text{deg}^2$ in the J band for kilonovae at 200 Mpc, but is limited to only $\approx 200$ $\text{deg}^2$ in the r band. At a distance of 100 (300) Mpc, these values change to $\approx 350$  (120) $\text{deg}^2$ for r band and 750 (220) $\text{deg}^2$ for J band. 

In Figure \ref{fig:time_distance_fraction}, we also plot the distances and areas  enclosing 90\% localization probabilities from the skymaps of the 15 events from one realization of our realistic BNS rate simulation. For an r-band search, seven of the 15 events have 90\% areas lying in the f$_{\rm{det}}>0.1$ region. In the J band, eight events lie in this region.

\subsection{WINTER GW follow-up strategy} \label{subsec:disc_obs}
WINTER is a dedicated instrument for follow-up of slowly fading infrared kilonovae and can spend weeks searching for a single event. However, even though longer searches cover larger fractions of the skymaps, they risk covering the high-probability areas less efficiently or observing the kilonova once it has already faded. 
Based on the results from simulating WINTER observations with the Kasen and Bulla models, WINTER will dedicate up to seven nights of searching for each event. WINTER will observe compelling transients until they fade, and if there are no candidate kilonovae in the data, will stop searching after seven nights. 

Furthermore, WINTER will not follow up all BNS GW alerts. Figure \ref{fig:time_distance_fraction} provides a prescription to select which GW triggers are worth following up with WINTER during O4 based on information that is available at the time of the trigger. We will only follow events that have median distance estimates and localization areas such that the chance of detecting a kilonova is at least 10\% (i.e., lying within the f$_{\rm{det}} > 0.1$ region of Figure \ref{fig:time_distance_fraction}). With a depth of $\text{J}_{AB} = 21$ (matching the WINTER reference images with a single exposure time t$_{\rm{exp}} = 450$ s), this means we can follow up events with distances of up to 350 Mpc if the time to tile the localization area is less than two days. We can follow up nearby events for much longer, with tiling times of six days for 200 Mpc. With WINTER's 1 $\text{deg}^2$ FOV, this corresponds to $\approx$ 150 $\text{deg}^2$ at 350 Mpc and 300 $\text{deg}^2$ at 250 Mpc. 

If the localization areas are larger and the events are nearer, we will reduce our exposure times to tile the localizations faster and increase the chances of detecting a kilonova. Figure \ref{fig:time_distances_depths} shows the $f_{\rm{det}} = 0.1$ contours for searches with depths of 21 mag (t$_{\rm{exp}} = 450$ s), 20.5 mag (t$_{\rm{exp}} = 180$ s) and 19.5 mag (t$_{\rm{exp}} = 40$ s) as a function of the localization area that can be searched with WINTER. With WINTER, we will follow up all events with localization areas $<300 $ $\text{deg}^2$ down to a depth of 21 magnitudes, matching WINTER's J-band reference images. For events with localization areas larger than 300 $\text{deg}^2$ and distances $< 250$ Mpc, we will reduce exposure time to 180 seconds (i.e. to a depth of 20.5 magnitudes). With 180 second exposures, we can tile areas as large as 1000 $\text{deg}^2$ while maintaining $f_{\rm{det}} = 0.1$. For events with localization areas larger than 1000 $\text{deg}^2$, we will follow only those events that are closer than 150 Mpc to a depth of $\text{J}_{AB} = 19.5$ magnitudes. If there are multiple gravitational-wave triggers of interest on the same night (as discussed in Table~\ref{tab:chip}), we will prioritize events that have a higher $f_{\rm{det}}$ and are easier to observe with WINTER.

Finally, we have assumed that the areas mentioned above enclose all of the BNS localization probability. If the full skymap areas are larger than the limits mentioned above, we will cover only those events where the area enclosing 50\% of the localization probability can be tiled with f$_{\rm{det}}>0.2$.
We also note that in Figure \ref{fig:time_distance_fraction}, the detection fraction f$_{\rm{det}}$ drops at very early search times ($t<0.5$ day), as the kilonova is still brightening in our models at these times. However, we will observe events with localization areas that can be tiled within 0.5 day, observing them repeatedly until the kilonova becomes bright enough to be detected. Similar detectability constraints were derived by ( \cite{Chase:2021ood}; their Figure 17) using the Los Alamos National Laboratory (LANL, \citealp{Wollaeger:2021qgf}) grid of kilonova models. The constraints presented in Figure \ref{fig:time_distance_fraction} are broadly consistent with their constraints, with minor variations attributable to the differences in the underlying Bulla and LANL models (see, for example, the differences in the analyses of \citet{Anand:2020eyg} and \citet{Thakur:2020yvu}, \citet{Dichiara:2021vjy}).

The methods described above can be used by other surveys to select GW triggers for follow-up during O4. We include a \texttt{python} notebook with the code to reproduce Figures 9--11 in the \texttt{Zenodo} repository at \cite{frostig_danielle_2021_5507322}.

\subsection{The realities of electromagnetic follow-up observing} \label{subsec:EM}

The methods outlined in Section \ref{subsec:gwemopt} represent a simplified approach to follow-up observations, where one observing strategy is decided at the outset and followed unchangingly throughout the campaign. In reality, follow-up observing is an iterative process where new decisions are made nightly or even multiple times per night. Instead of solely following one search strategy as shown in the above simulations, we will prioritize repeat observations of compelling transients found in the follow-up data from WINTER and other telescopes. In simulations, WINTER observes at the zenith to a limiting magnitude of $\text{J}_{AB} = 21$ in 450 seconds, $\text{J}_{AB} = 21.8$ in 30 minutes, and $\text{J}_{AB} = 22.7$ in 3 hours, with many repeat observations allowing for more significant constraints on kilonova detection. 

Furthermore, if another telescope discovers a new kilonova candidate, infrared localizations, not just discoveries, add unique data to kilonova science. A nondetection in WINTER of an optical kilonova constrains both the dynamical and wind ejecta masses \citep{Chase:2021ood}. Additionally, faint detections ($\text{SNR}_{\text{EM}}<5$) or single images of the event in the near infrared contribute to models of chemical evolution, ejecta mass, and wind speeds \citep{Barnes:2016umi, Coughlin:2019zqi, Kasen:2013xka}.

However, with no confirmed discoveries from other telescopes, there can be thousands of candidate transient events to sort through in a week of near infrared survey data. There are many tools available for classifying transients in survey data with machine-learning algorithms becoming a standard tool in the field, particularly for wide-field optical surveys, such as ZTF and the Vera C. Rubin Observatory Legacy Survey of Space and Time (LSST; \cite{Forster:2020pxc, G_mez_2020, DES:2015sji, Mahabal_2019,  Sooknunan_2020, Turpin:2020dti}).

Carefully planning the color and cadence of WINTER follow-up observations can assist these transient classification techniques in narrowing down the number of candidate events. For example, the characteristic $\sim 1$ week fading time distinguishes a kilonova lightcurve from longer-lasting supernovae or short-lived asteroids \citep{Cowperthwaite:2015kya}. Additionally, observations in at least two filters assist in studying the reddening of the transient over time, a distinctive feature of kilonovae seen in GW170817 \citep{Arcavi:2017xiz, Pian:2017gtc, Smartt:2017fuw}. For optical telescopes, the i band provides the reddest images, and models predict kilonova discovery is maximized with the g –- i filter pair for LSST \citep{LSSTTransient:2018dvm} and a g, r, i filter cycle for ZTF \citep{Almualla:2020ybm}. In the simulation described in Section~\ref{subsec:gwemopt}, we only observe in the J-band to leverage WINTER's J-band reference images and all-sky survey. In practice, we aim to conduct most of the search in J band but also follow up all interesting candidate events in the Y-band to study the Y –- J color evolution. We will also study the g –- J color pair, either with g-band images from ZTF follow-up or with the optical camera on WINTER's companion port. Time permitting, WINTER and its counterpart optical camera will also observe candidate events in the u, r, i, and H filters. 

\section{Conclusion} \label{sec:summary}
The BNS merger GW170817 brought about a new field of multimessenger astronomy, but despite extensive follow-up campaigns in O3, we have not observed a second multimessenger kilonova. In this study, we show infrared observations are a promising avenue for kilonova discovery, particularly for lanthanide-rich ``red" kilonovae, as these are detectable to larger distances in the infrared than at optical wavelengths. We predict that infrared follow-up of GW triggers with WINTER could discover up to ten new kilonovae per year during O4. Furthermore, by employing more targeted follow-up strategies than those we have simulated, we can achieve a deeper sensitivity on a subset of interesting targets, therefore enhancing our ability to confirm new discoveries. 

Moreover, we limit this study to BNS mergers and leave the study of NSBH kilonovae to a future work. Infrared follow-up of NSBH kilonovae is especially promising, as they are brighter in the infrared compared to BNS kilonovae \citep{Anand:2020eyg, Fernandez:2016sbf, Zhu:2020inc}, and observing both event types has the potential to increase the number of kilonovae discovered each year. Even just one new electromagnetic observation of a kilonova in O4 will double the number of known multimessenger kilonovae, helping to answer ongoing questions in the study of the Hubble tension, the neutron star equation of state, and r-process nucleosynthesis.

\software{\texttt{astropy} \citep{astropy:2013},
         \gwemopt \citep{Coughlin:2018lta}, 
         \texttt{gwemlightcurves} \citep{Coughlin:2018miv, Coughlin:2018fis, Dietrich:2020efo}, 
         \texttt{matplotlib} \citep{Hunter:2007},
         \texttt{numpy} \citep{harris2020array},
         \texttt{ligo.skymap} \citep{Singer:2015ema, Singer:2016eax, Singer:2016erz},
         \texttt{pandas} \citep{mckinney2010data}, 
         \bilby~\citep{Ashton:2018jfp, Romero-Shaw:2020owr}, 
         \pycbc~\citep{Nitz:2018rgo, DalCanton:2020vpm}, 
         \texttt{PyMultinest}~\citep{Feroz:2007kg, Feroz:2008xx, Feroz:2013hea, Buchner:2014nha}
         }

\acknowledgments
The authors thank Michael Coughlin for his support integrating WINTER into \gwemopt and for help with \gwemlightcurves.
WINTER's construction is made possible by the National Science Foundation under MRI grant number AST-1828470.  We also acknowledge significant support for WINTER from the California Institute of Technology, the Caltech Optical Observatories, the Bruno Rossi Fund of the MIT Kavli Institute for Astrophysics and Space Research, and the MIT Department of Physics and School of Science.
S.~B., G.~M., H.-Y.~C., E.~K. and S.~V\ acknowledge support of the National Science Foundation and the LIGO Laboratory.
LIGO was constructed by the California Institute of Technology and
Massachusetts Institute of Technology with funding from the National
Science Foundation and operates under cooperative agreement PHY-0757058.
S.~V. is supported by the NSF through award PHY-2045740.
S.~B. is also supported by the NSF Graduate Research Fellowship under grant No. DGE-1122374.
G.~M. is supported by the NSF through award PHY-1764464.
M.~M.~K. acknowledges generous support from the David and Lucille Packard Foundation.
The authors are grateful for computational resources provided by the LIGO Lab and supported by NSF Grants PHY-0757058 and PHY-0823459.
This paper carries LIGO document number LIGO-P2100340.

\bibliography{ligowinter}{}

\begin{thebibliography}{}
\expandafter\ifx\csname natexlab\endcsname\relax\def\natexlab#1{#1}\fi
\providecommand{\url}[1]{\href{#1}{#1}}

\bibitem[{Abb(2017)}]{Abbott:2017xzu}
 2017, Nature, 551, 85–88.
\newblock \url{http://dx.doi.org/10.1038/nature24471}

\bibitem[{Aasi {et~al.}(2015)Aasi, Abbott, Abbott, Abbott, Abernathy, Ackley,
  Adams, Adams, Addesso, \& et~al.}]{TheLIGOScientific:2014jea}
Aasi, J., Abbott, B.~P., Abbott, R., {et~al.} 2015, Class. Quant. Grav., 32,
  074001.
\newblock \url{http://dx.doi.org/10.1088/0264-9381/32/7/074001}

\bibitem[{Abbott {et~al.}(2017{\natexlab{a}})Abbott, Abbott, Abbott, Acernese,
  Ackley, Adams, Adams, Addesso, Adhikari, Adya, \&
  et~al.}]{TheLIGOScientific:2017qsa}
Abbott, B., Abbott, R., Abbott, T., {et~al.} 2017{\natexlab{a}}, Physical
  Review Letters, 119, doi:10.1103/physrevlett.119.161101.
\newblock \url{http://dx.doi.org/10.1103/PhysRevLett.119.161101}

\bibitem[{Abbott {et~al.}(2017{\natexlab{b}})Abbott, Abbott, Abbott, Acernese,
  Ackley, Adams, Adams, Addesso, Adhikari, Adya, \&
  et~al.}]{LIGOScientific:2017vwq}
---. 2017{\natexlab{b}}, Physical Review Letters, 119,
  doi:10.1103/physrevlett.119.161101.
\newblock \url{http://dx.doi.org/10.1103/PhysRevLett.119.161101}

\bibitem[{Abbott {et~al.}(2020{\natexlab{a}})Abbott, Abbott, Abbott, \&
  et~al.}]{noise_curves}
Abbott, B.~P., Abbott, R., Abbott, T.~D., \& et~al. 2020{\natexlab{a}}, Noise
  curves used for Simulations in the update of the Observing Scenarios Paper,
  \url{https://dcc.ligo.org/LIGO-T2000012/public}, ,

\bibitem[{Abbott {et~al.}(2017{\natexlab{c}})Abbott, Abbott, Abbott, Acernese,
  Ackley, Adams, Adams, Addesso, Adhikari, Adya, \& et~al.}]{GBM:2017lvd}
Abbott, B.~P., Abbott, R., Abbott, T.~D., {et~al.} 2017{\natexlab{c}}, The
  Astrophysical Journal, 848, L12.
\newblock \url{http://dx.doi.org/10.3847/2041-8213/aa91c9}

\bibitem[{Abbott {et~al.}(2020{\natexlab{b}})Abbott, Abbott, Abbott, Abraham,
  Acernese, Ackley, Adams, Adya, Affeldt, Agathos, \& et~al.}]{Abbott:2020qfu}
---. 2020{\natexlab{b}}, Living Rev. Rel., 23, 3.
\newblock \url{http://dx.doi.org/10.1007/s41114-020-00026-9}

\bibitem[{Abbott {et~al.}(2020{\natexlab{c}})Abbott, Abbott, Abbott, Abraham,
  Acernese, Ackley, Adams, Adya, Affeldt, Agathos, \& et~al.}]{Aasi:2013wya}
---. 2020{\natexlab{c}}, Living Reviews in Relativity, 23,
  doi:10.1007/s41114-020-00026-9.
\newblock \url{http://dx.doi.org/10.1007/s41114-020-00026-9}

\bibitem[{Abbott {et~al.}(2021{\natexlab{a}})Abbott, Abbott, Abraham, Acernese,
  Ackley, Adams, Adams, Adhikari, Adya, Affeldt, \&
  et~al.}]{LIGOScientific:2021qlt}
Abbott, R., Abbott, T.~D., Abraham, S., {et~al.} 2021{\natexlab{a}}, The
  Astrophysical Journal Letters, 915, L5.
\newblock \url{http://dx.doi.org/10.3847/2041-8213/ac082e}

\bibitem[{Abbott {et~al.}(2021{\natexlab{b}})Abbott, Abbott, Abraham, Acernese,
  Ackley, Adams, Adams, Adhikari, Adya, Affeldt, \& et~al.}]{Abbott:2020niy}
Abbott, R., Abbott, T., Abraham, S., {et~al.} 2021{\natexlab{b}}, Physical
  Review X, 11, doi:10.1103/physrevx.11.021053.
\newblock \url{http://dx.doi.org/10.1103/PhysRevX.11.021053}

\bibitem[{Acernese {et~al.}(2014)Acernese, Agathos, Agatsuma, Aisa, Allemandou,
  Allocca, Amarni, Astone, Balestri, Ballardin, \& et~al.}]{TheVirgo:2014hva}
Acernese, F., Agathos, M., Agatsuma, K., {et~al.} 2014, Classical and Quantum
  Gravity, 32, 024001.
\newblock \url{http://dx.doi.org/10.1088/0264-9381/32/2/024001}

\bibitem[{Ackley {et~al.}(2020)Ackley, Amati, Barbieri, Bauer, Benetti,
  Bernardini, Bhirombhakdi, Botticella, Branchesi, Brocato, \&
  et~al.}]{Ackley:2020qkz}
Ackley, K., Amati, L., Barbieri, C., {et~al.} 2020, Astronomy \& Astrophysics,
  643, A113.
\newblock \url{http://dx.doi.org/10.1051/0004-6361/202037669}

\bibitem[{Akmal \& Pandharipande(1997)}]{Akmal:1997ft}
Akmal, A., \& Pandharipande, V.~R. 1997, Phys. Rev. C, 56, 2261

\bibitem[{Akmal {et~al.}(1998)Akmal, Pandharipande, \&
  Ravenhall}]{Akmal:1998cf}
Akmal, A., Pandharipande, V.~R., \& Ravenhall, D.~G. 1998, Phys. Rev. C, 58,
  1804

\bibitem[{Akutsu {et~al.}(2018)Akutsu, Ando, Araki, Araya, Arima, Aritomi,
  Asada, Aso, Atsuta, Awai, \& et~al.}]{Akutsu:2017kpk}
Akutsu, T., Ando, M., Araki, S., {et~al.} 2018, Progress of Theoretical and
  Experimental Physics, 2018, doi:10.1093/ptep/ptx180.
\newblock \url{http://dx.doi.org/10.1093/ptep/ptx180}

\bibitem[{Allen {et~al.}(2012)Allen, Anderson, Brady, Brown, \&
  Creighton}]{Allen:2005fk}
Allen, B., Anderson, W.~G., Brady, P.~R., Brown, D.~A., \& Creighton, J. D.~E.
  2012, Phys. Rev. D, 85, 122006

\bibitem[{Almualla {et~al.}(2021)Almualla, Anand, Coughlin, Dietrich, Guessoum,
  Sagués~Carracedo, Ahumada, Andreoni, Antier, Bellm, \&
  et~al.}]{Almualla:2020ybm}
Almualla, M., Anand, S., Coughlin, M.~W., {et~al.} 2021, Monthly Notices of the
  Royal Astronomical Society, 504, 2822–2831.
\newblock \url{http://dx.doi.org/10.1093/mnras/stab1090}

\bibitem[{{Alpar} {et~al.}(1982){Alpar}, {Cheng}, {Ruderman}, \&
  {Shaham}}]{1982Natur.300..728A}
{Alpar}, M.~A., {Cheng}, A.~F., {Ruderman}, M.~A., \& {Shaham}, J. 1982, \nat,
  300, 728

\bibitem[{Anand {et~al.}(2020)Anand, Coughlin, Kasliwal, Bulla, Ahumada,
  Sagués~Carracedo, Almualla, Andreoni, Stein, Foucart, \&
  et~al.}]{Anand:2020eyg}
Anand, S., Coughlin, M.~W., Kasliwal, M.~M., {et~al.} 2020, Nature Astronomy,
  5, 46–53.
\newblock \url{http://dx.doi.org/10.1038/s41550-020-1183-3}

\bibitem[{Andreoni {et~al.}(2019)Andreoni, Anand, Bianco, Cenko, Cowperthwaite,
  Coughlin, Drout, Golkhou, Kaplan, Mooley, \& et~al.}]{LSSTTransient:2018dvm}
Andreoni, I., Anand, S., Bianco, F.~B., {et~al.} 2019, Publications of the
  Astronomical Society of the Pacific, 131, 068004.
\newblock \url{http://dx.doi.org/10.1088/1538-3873/ab1531}

\bibitem[{Andreoni {et~al.}(2020)Andreoni, Kool, Carracedo, Kasliwal, Bulla,
  Ahumada, Coughlin, Anand, Sollerman, Goobar, \& et~al.}]{Andreoni:2020ewy}
Andreoni, I., Kool, E.~C., Carracedo, A.~S., {et~al.} 2020, The Astrophysical
  Journal, 904, 155.
\newblock \url{http://dx.doi.org/10.3847/1538-4357/abbf4c}

\bibitem[{Antier {et~al.}(2020)Antier, Agayeva, Almualla, Awiphan, Baransky,
  Barynova, Beradze, Blažek, Boër, Burkhonov, \& et~al.}]{Antier:2020nuy}
Antier, S., Agayeva, S., Almualla, M., {et~al.} 2020, Monthly Notices of the
  Royal Astronomical Society, 497, 5518–5539.
\newblock \url{http://dx.doi.org/10.1093/mnras/staa1846}

\bibitem[{Arcavi {et~al.}(2017)Arcavi, Hosseinzadeh, Howell, McCully,
  Poznanski, Kasen, Barnes, Zaltzman, Vasylyev, Maoz, \&
  et~al.}]{Arcavi:2017xiz}
Arcavi, I., Hosseinzadeh, G., Howell, D.~A., {et~al.} 2017, Nature, 551,
  64–66.
\newblock \url{http://dx.doi.org/10.1038/nature24291}

\bibitem[{Ashton {et~al.}(2019)Ashton, Hübner, Lasky, Talbot, Ackley,
  Biscoveanu, Chu, Divakarla, Easter, Goncharov, \& et~al.}]{Ashton:2018jfp}
Ashton, G., Hübner, M., Lasky, P.~D., {et~al.} 2019, The Astrophysical Journal
  Supplement Series, 241, 27.
\newblock \url{http://dx.doi.org/10.3847/1538-4365/ab06fc}

\bibitem[{Aso {et~al.}(2013)Aso, Michimura, Somiya, Ando, Miyakawa, Sekiguchi,
  Tatsumi, \& Yamamoto}]{Aso:2013eba}
Aso, Y., Michimura, Y., Somiya, K., {et~al.} 2013, Phys. Rev. D, 88, 043007

\bibitem[{Babak(2008)}]{Babak:2008rb}
Babak, S. 2008, Class. Quant. Grav., 25, 195011

\bibitem[{Barnes \& Kasen(2013)}]{Barnes:2013wka}
Barnes, J., \& Kasen, D. 2013, Astrophys. J., 775, 18

\bibitem[{Barnes {et~al.}(2016)Barnes, Kasen, Wu, \&
  Mart\'\i{}nez-Pinedo}]{Barnes:2016umi}
Barnes, J., Kasen, D., Wu, M.-R., \& Mart\'\i{}nez-Pinedo, G. 2016, Astrophys.
  J., 829, 110

\bibitem[{Bauswein {et~al.}(2013)Bauswein, Baumgarte, \&
  Janka}]{Bauswein:2013jpa}
Bauswein, A., Baumgarte, T.~W., \& Janka, H.~T. 2013, Phys. Rev. Lett., 111,
  131101

\bibitem[{{Becerra} {et~al.}(2021){Becerra}, {Dichiara}, {Watson}, {Troja},
  {Butler}, {Pereyra}, {Moreno M{\'e}ndez}, {De Colle}, {Lee}, {Kutyrev}, \&
  {L{\'o}pez}}]{2021MNRAS.507.1401B}
{Becerra}, R.~L., {Dichiara}, S., {Watson}, A.~M., {et~al.} 2021, \mnras, 507,
  1401

\bibitem[{Bellm {et~al.}(2018)Bellm, Kulkarni, Graham, Dekany, Smith, Riddle,
  Masci, Helou, Prince, Adams, \& et~al.}]{Bellm_2018}
Bellm, E.~C., Kulkarni, S.~R., Graham, M.~J., {et~al.} 2018, Publications of
  the Astronomical Society of the Pacific, 131, 018002.
\newblock \url{http://dx.doi.org/10.1088/1538-3873/aaecbe}

\bibitem[{Biscoveanu {et~al.}(2019)Biscoveanu, Vitale, \&
  Haster}]{Biscoveanu:2019ugx}
Biscoveanu, S., Vitale, S., \& Haster, C.-J. 2019, Astrophys. J. Lett., 884,
  L32

\bibitem[{Breschi {et~al.}(2021)Breschi, Perego, Bernuzzi, Del~Pozzo, Nedora,
  Radice, \& Vescovi}]{Breschi:2021tbm}
Breschi, M., Perego, A., Bernuzzi, S., {et~al.} 2021, Monthly Notices of the
  Royal Astronomical Society, 505, 1661–1677.
\newblock \url{http://dx.doi.org/10.1093/mnras/stab1287}

\bibitem[{Buchner {et~al.}(2014)Buchner, Georgakakis, Nandra, Hsu, Rangel,
  Brightman, Merloni, Salvato, Donley, \& Kocevski}]{Buchner:2014nha}
Buchner, J., Georgakakis, A., Nandra, K., {et~al.} 2014, Astron. Astrophys.,
  564, A125

\bibitem[{Bulla(2019)}]{Bulla:2019muo}
Bulla, M. 2019, Mon. Not. Roy. Astron. Soc., 489, 5037

\bibitem[{Burgay {et~al.}(2003)Burgay, D’Amico, Possenti, Manchester, Lyne,
  Joshi, McLaughlin, Kramer, Sarkissian, Camilo, \& et~al.}]{Burgay:2003jj}
Burgay, M., D’Amico, N., Possenti, A., {et~al.} 2003, Nature, 426, 531–533.
\newblock \url{http://dx.doi.org/10.1038/nature02124}

\bibitem[{{Catelan} {et~al.}(2011){Catelan}, {Minniti}, {Lucas},
  {Alonso-Garc{\'\i}a}, {Angeloni}, {Beam{\'\i}n}, {Bonatto}, {Borissova},
  {Contreras}, {Cross}, {D{\'e}k{\'a}any}, {Emerson}, {Eyheramendy}, {Geisler},
  {Gonz{\'a}lez-Solares}, {Helminiak}, {Hempel}, {Irwin}, {Ivanov},
  {Jord{\'a}n}, {Kerins}, {Kurtev}, {Mauro}, {Moni Bidin}, {Navarrete},
  {P{\'e}rez}, {Pichara}, {Read}, {Rejkuba}, {Saito}, {Sale}, \&
  {Toledo}}]{catelan2011vista}
{Catelan}, M., {Minniti}, D., {Lucas}, P.~W., {et~al.} 2011, {The Vista
  Variables in the V{\'\i}a L{\'a}ctea (VVV) ESO Public Survey: Current Status
  and First Results}, , , arXiv:1105.1119

\bibitem[{Chase {et~al.}(2021)Chase, O'Connor, Fryer, Troja, Korobkin,
  Wollaeger, Ristic, Fontes, Hungerford, \& Herring}]{Chase:2021ood}
Chase, E.~A., O'Connor, B., Fryer, C.~L., {et~al.} 2021, arXiv:2105.12268

\bibitem[{Chen {et~al.}(2018)Chen, Fishbach, \& Holz}]{Chen:2017rfc}
Chen, H.-Y., Fishbach, M., \& Holz, D.~E. 2018, Nature, 562, 545

\bibitem[{Collaboration {et~al.}(2021)Collaboration, the Virgo~Collaboration,
  Abbott, Abbott, Acernese, Ackley, Adams, Adhikari, Adhikari, Adya, Affeldt,
  Agarwal, Agathos, Agatsuma, Aggarwal, Aguiar, Aiello, Ain, Ajith, Albanesi,
  Allocca, Altin, Amato, Anand, Anand, Ananyeva, Anderson, Anderson, Andrade,
  Andres, Andrić, Angelova, Ansoldi, Antelis, Antier, Appert, Arai, Araya,
  Areeda, Arène, Arnaud, Aronson, Arun, Asali, Ashton, Assiduo, Aston, Astone,
  Aubin, Austin, Babak, Badaracco, Bader, Badger, Bae, Baer, Bagnasco, Bai,
  Baird, Ball, Ballardin, Ballmer, Balsamo, Baltus, Banagiri, Bankar, Barayoga,
  Barbieri, Barish, Barker, Barneo, Barone, Barr, Barsotti, Barsuglia, Barta,
  Bartlett, Barton, Bartos, Bassiri, Basti, Bawaj, Bayley, Baylor, Bazzan,
  Bécsy, Bedakihale, Bejger, Belahcene, Benedetto, Beniwal, Bennett, Bentley,
  BenYaala, Bergamin, Berger, Bernuzzi, Berry, Bersanetti, Bertolini,
  Betzwieser, Beveridge, Bhandare, Bhardwaj, Bhattacharjee, Bhaumik, Bilenko,
  Billingsley, Bini, Birney, Birnholtz, Biscans, Bischi, Biscoveanu, Bisht,
  Biswas, Bitossi, Bizouard, Blackburn, Blair, Blair, Blair, Bobba, Bode, Boer,
  Bogaert, Boldrini, Bonavena, Bondu, Bonilla, Bonnand, Booker, Boom, Bork,
  Boschi, Bose, Bose, Bossilkov, Boudart, Bouffanais, Bozzi, Bradaschia, Brady,
  Bramley, Branch, Branchesi, Brau, Breschi, Briant, Briggs, Brillet,
  Brinkmann, Brockill, Brooks, Brooks, Brown, Brunett, Bruno, Bruntz, Bryant,
  Bulik, Bulten, Buonanno, Buscicchio, Buskulic, Buy, Byer, Cadonati, Cagnoli,
  Cahillane, Bustillo, Callaghan, Callister, Calloni, Cameron, Camp, Canepa,
  Canevarolo, Cannavacciuolo, Cannon, Cao, Capote, Carapella, Carbognani,
  Carlin, Carney, Carpinelli, Carrillo, Carullo, Carver, Diaz, Casentini,
  Castaldi, Caudill, Cavaglià, Cavalier, Cavalieri, Ceasar, Cella,
  Cerdá-Durán, Cesarini, Chaibi, Chakravarti, Subrahmanya, Champion, Chan,
  Chan, Chan, Chan, Chandra, Chanial, Chao, Charlton, Chase, Chassande-Mottin,
  Chatterjee, Chatterjee, Chatterjee, Chattopadhyay, Chaturvedi, Chaty,
  Chatziioannou, Chen, Chen, Chen, Chen, Chen, Cheng, Cheong, Cheung, Chia,
  Chiadini, Chiarini, Chierici, Chincarini, Chiofalo, Chiummo, Cho, Cho,
  Choudhary, Choudhary, Christensen, Chu, Chua, Chung, Ciani, Ciecielag,
  Cieślar, Cifaldi, Ciobanu, Ciolfi, Cipriano, Cirone, Clara, Clark, Clark,
  Clarke, Clearwater, Clesse, Cleva, Coccia, Codazzo, Cohadon, Cohen, Cohen,
  Colleoni, Collette, Colombo, Colpi, Compton, au2, Conti, Cooper, Corban,
  Corbitt, Cordero-Carrión, Corezzi, Corley, Cornish, Corre, Corsi, Cortese,
  Costa, Cotesta, Coughlin, Coulon, Countryman, Cousins, Couvares, Coward,
  Cowart, Coyne, Coyne, Creighton, Creighton, Criswell, Croquette, Crowder,
  Cudell, Cullen, Cumming, Cummings, Cunningham, Cuoco, Curyło, Dabadie,
  Canton, Dall'Osso, Dálya, Dana, DaneshgaranBajastani, D'Angelo, Danila,
  Danilishin, D'Antonio, Danzmann, Darsow-Fromm, Dasgupta, Datrier, Datta,
  Dattilo, Dave, Davier, Davies, Davis, Davis, Daw, Dean, DeBra, Deenadayalan,
  Degallaix, Laurentis, Deléglise, Favero, Lillo, Lillo, Pozzo, DeMarchi,
  Matteis, D'Emilio, Demos, Dent, Depasse, Pietri, Rosa, Rossi, DeSalvo,
  Simone, Dhurandhar, Díaz, au2, Didio, Dietrich, Fiore, Fronzo, Giorgio,
  Giovanni, Giovanni, Girolamo, Lieto, Ding, Pace, Palma, Renzo, Divakarla,
  Divyajyoti, Dmitriev, Doctor, D'Onofrio, Donovan, Dooley, Doravari,
  Dorrington, Drago, Driggers, Drori, Ducoin, Dupej, Durante, D'Urso, Duverne,
  Dwyer, Eassa, Easter, Ebersold, Eckhardt, Eddolls, Edelman, Edo, Edy, Effler,
  Eichholz, Eikenberry, Eisenmann, Eisenstein, Ejlli, Engelby, Errico, Essick,
  Estellés, Estevez, Etienne, Etzel, Evans, Evans, Ewing, Fafone, Fair,
  Fairhurst, Fanning, Farah, Farinon, Farr, Farr, Farrow, Fauchon-Jones,
  Favaro, Favata, Fays, Fazio, Feicht, Fejer, Fenyvesi, Ferguson,
  Fernandez-Galiana, Ferrante, Ferreira, Fidecaro, Figura, Fiori, Fishbach,
  Fisher, Fittipaldi, Fiumara, Flaminio, Floden, Fong, Font, Fornal, Forsyth,
  Franke, Frasca, Frasconi, Frederick, Freed, Frei, Freise, Frey, Fritschel,
  Frolov, Fronzé, Fulda, Fyffe, Gabbard, Gabella, Gadre, Gair, Gais,
  Galaudage, Gamba, Ganapathy, Ganguly, Gaonkar, Garaventa, García,
  García-Núñez, García-Quirós, Garufi, Gateley, Gaudio, Gayathri, Gemme,
  Gennai, George, George, Gerberding, Gergely, Gewecke, Ghonge, Ghosh, Ghosh,
  Ghosh, Ghosh, Giacomazzo, Giacoppo, Giaime, Giardina, Gibson, Gier, Giesler,
  Giri, Gissi, Glanzer, Gleckl, Godwin, Goetz, Goetz, Gohlke, Goncharov,
  González, Gopakumar, Gosselin, Gouaty, Gould, Grace, Grado, Granata,
  Granata, Grant, Gras, Grassia, Gray, Gray, Greco, Green, Green, Gretarsson,
  Gretarsson, Griffith, Griffiths, Griggs, Grignani, Grimaldi, Grimm, Grote,
  Grunewald, Gruning, Guerra, Guidi, Guimaraes, Guixé, Gulati, Guo, Guo,
  Gupta, Gupta, Gupta, Gustafson, Gustafson, Guzman, Haegel, Halim, Hall,
  Hamilton, Hammond, Haney, Hanks, Hanna, Hannam, Hannuksela, Hansen, Hansen,
  Hanson, Harder, Hardwick, Haris, Harms, Harry, Harry, Hartwig, Haskell,
  Hasskew, Haster, Haughian, Hayes, Healy, Heidmann, Heidt, Heintze, Heinze,
  Heinzel, Heitmann, Hellman, Hello, Helmling-Cornell, Hemming, Hendry, Heng,
  Hennes, Hennig, Hennig, Hernandez, Vivanco, Heurs, Hild, Hill, Hines,
  Hochheim, Hofman, Hohmann, Holcomb, Holland, Holley-Bockelmann, Hollows,
  Holmes, Holt, Holz, Hopkins, Hough, Hourihane, Howell, Hoy, Hoyland, Hreibi,
  Hsu, Huang, Hübner, Huddart, Hughey, Hui, Husa, Huttner, Huxford,
  Huynh-Dinh, Idzkowski, Iess, Ingram, Isi, Isleif, Iyer, JaberianHamedan,
  Jacqmin, Jadhav, Jadhav, James, Jan, Jani, Janquart, Janssens, Janthalur,
  Jaranowski, Jariwala, Jaume, Jenkins, Jenner, Jeunon, Jia, Johns,
  Johnson-McDaniel, Jones, Jones, Jones, Jones, Jones, Jonker, Ju, Junker,
  Juste, Kalaghatgi, Kalogera, Kamai, Kandhasamy, Kang, Kanner, Kao, Kapadia,
  Kapasi, Karat, Karathanasis, Karki, Kashyap, Kasprzack, Kastaun, Katsanevas,
  Katsavounidis, Katzman, Kaur, Kawabe, Kéfélian, Keitel, Key, Khadka,
  Khalili, Khan, Khazanov, Khetan, Khursheed, Kijbunchoo, Kim, Kim, Kim, Kim,
  Kim, Kimball, Kinley-Hanlon, Kirchhoff, Kissel, Kleybolte, Klimenko, Knee,
  Knowles, Knyazev, Koch, Koekoek, Koley, Kolitsidou, Kolstein, Komori,
  Kondrashov, Kontos, Koper, Korobko, Kovalam, Kozak, Kringel, Krishnendu,
  Królak, Kuehn, Kuei, Kuijer, Kumar, Kumar, Kumar, Kumar, Kuns, Kuwahara,
  Lagabbe, Laghi, Lalande, Lam, Lamberts, Landry, Lane, Lang, Lange, Lantz,
  Rosa, Lartaux-Vollard, Lasky, Laxen, Lazzarini, Lazzaro, Leaci, Leavey,
  Lecoeuche, Lee, Lee, Lee, Lee, Lehmann, Lemaître, Leroy, Letendre, Levesque,
  Levin, Leviton, Leyde, Li, Li, Li, Li, Li, Linde, Linker, Linley, Littenberg,
  Liu, Liu, Liu, Llamas, Llorens-Monteagudo, Lo, Lockwood, London, Longo,
  Lopez, Portilla, Lorenzini, Loriette, Lormand, Losurdo, Lott, Lough, Lousto,
  Lovelace, Lucaccioni, Lück, Lumaca, Lundgren, Lynam, Macas, MacInnis,
  Macleod, MacMillan, Macquet, Hernandez, Magazzù, Magee, Maggiore, Magnozzi,
  Mahesh, Majorana, Makarem, Maksimovic, Maliakal, Malik, Man, Mandic, Mangano,
  Mango, Mansell, Manske, Mantovani, Mapelli, Marchesoni, Marion, Mark, Márka,
  Márka, Markakis, Markosyan, Markowitz, Maros, Marquina, Marsat, Martelli,
  Martin, Martin, Martinez, Martinez, Martinez, Martinovic, Martynov, Marx,
  Masalehdan, Mason, Massera, Masserot, Massinger, Masso-Reid, Mastrogiovanni,
  Matas, Mateu-Lucena, Matichard, Matiushechkina, Mavalvala, McCann, McCarthy,
  McClelland, McClincy, McCormick, McCuller, McGhee, McGuire, McIsaac, McIver,
  McRae, McWilliams, Meacher, Mehmet, Mehta, Meijer, Melatos, Melchor, Mendell,
  Menendez-Vazquez, Menoni, Mercer, Mereni, Merfeld, Merilh, Merritt,
  Merzougui, Meshkov, Messenger, Messick, Meyers, Meylahn, Mhaske, Miani, Miao,
  Michaloliakos, Michel, Middleton, Milano, Miller, Miller, Miller, Millhouse,
  Mills, Milotti, Minazzoli, Minenkov, Mir, Miravet-Tenés, Mishra, Mishra,
  Mistry, Mitra, Mitrofanov, Mitselmakher, Mittleman, Mo, Moguel, Mogushi,
  Mohapatra, Mohite, Molina, Molina-Ruiz, Mondin, Montani, Moore, Moraru,
  Morawski, More, Moreno, Moreno, Morisaki, Mours, Mow-Lowry, Mozzon,
  Muciaccia, Mukherjee, Mukherjee, Mukherjee, Mukherjee, Mukherjee, Mukund,
  Mullavey, Munch, Muñiz, Murray, Musenich, Muusse, Nadji, Nagar, Napolano,
  Nardecchia, Naticchioni, Nayak, Nayak, Neil, Neilson, Nelemans, Nelson, Nery,
  Neubauer, Neunzert, Ng, Ng, Nguyen, Nguyen, Nguyen, Nichols, Nissanke,
  Nitoglia, Nocera, Norman, North, Nuttall, Oberling, O'Brien, O'Dell, Oelker,
  Oganesyan, Oh, Oh, Ohme, Ohta, Okada, Olivetto, Oram, O'Reilly, Ormiston,
  Ormsby, Ortega, O'Shaughnessy, O'Shea, Ossokine, Osthelder, Ottaway,
  Overmier, Pace, Pagano, Page, Pagliaroli, Pai, Pai, Palamos, Palashov,
  Palomba, Pan, Panda, Pang, Pankow, Pannarale, Pant, Panther, Paoletti, Paoli,
  Paolone, Park, Parker, Pascucci, Pasqualetti, Passaquieti, Passuello, Patel,
  Pathak, Patricelli, Patron, Patrone, Paul, Payne, Pedraza, Pegoraro, Pele,
  Penn, Perego, Pereira, Pereira, Perez, Périgois, Perkins, Perreca, Perriès,
  Petermann, Petterson, Pfeiffer, Pham, Phukon, Piccinni, Pichot, Piendibene,
  Piergiovanni, Pierini, Pierro, Pillant, Pillas, Pilo, Pinard, Pinto, Pinto,
  Piotrzkowski, Pirello, Pitkin, Placidi, Planas, Plastino, Pluchar, Poggiani,
  Polini, Pong, Ponrathnam, Popolizio, Porter, Poulton, Powell, Pracchia,
  Pradier, Prajapati, Prasai, Prasanna, Pratten, Principe, Prodi, Prokhorov,
  Prosposito, Prudenzi, Puecher, Punturo, Puosi, Puppo, Pürrer, Qi, Quetschke,
  Quitzow-James, Raab, Raaijmakers, Radkins, Radulesco, Raffai, Rail, Raja,
  Rajan, Ramirez, Ramirez, Ramos-Buades, Rana, Rapagnani, Rapol, Ray, Raymond,
  Raza, Razzano, Read, Rees, Regimbau, Rei, Reid, Reid, Reitze, Relton,
  Renzini, Rettegno, Reza, Rezac, Ricci, Richards, Richardson, Richardson,
  Riemenschneider, Riles, Rinaldi, Rink, Rizzo, Robertson, Robie, Robinet,
  Rocchi, Rodriguez, Rolland, Rollins, Romanelli, Romano, Romel,
  Romero-Rodríguez, Romero-Shaw, Romie, Ronchini, Rosa, Rose, Rosell,
  Rosińska, Ross, Rowan, Rowlinson, Roy, Roy, Roy, Rozza, Ruggi, Ruiz-Rocha,
  Ryan, Sachdev, Sadecki, Sadiq, Sakellariadou, Salafia, Salconi, Saleem,
  Salemi, Samajdar, Sanchez, Sanchez, Sanchez, Sanchis-Gual, Sanders, Sanuy,
  Saravanan, Sarin, Sassolas, Satari, Sauter, Savage, Sawant, Sawant, Sayah,
  Schaetzl, Scheel, Scheuer, Schiworski, Schmidt, Schmidt, Schnabel,
  Schneewind, Schofield, Schönbeck, Schulte, Schutz, Schwartz, Scott, Scott,
  Seglar-Arroyo, Sellers, Sengupta, Sentenac, Seo, Sequino, Sergeev, Setyawati,
  Shaffer, Shahriar, Shams, Sharma, Sharma, Shawhan, Shcheblanov, Shikauchi,
  Shoemaker, Shoemaker, ShyamSundar, Sieniawska, Sigg, Singer, Singh, Singh,
  Singha, Sintes, Sipala, Skliris, Slagmolen, Slaven-Blair, Smetana, Smith,
  Smith, Soldateschi, Somala, Son, Soni, Soni, Sordini, Sorrentino, Sorrentino,
  Soulard, Souradeep, Sowell, Spagnuolo, Spencer, Spera, Srinivasan,
  Srivastava, Srivastava, Staats, Stachie, Steer, Steinhoff, Steinlechner,
  Steinlechner, Stevenson, Stops, Stover, Strain, Strang, Stratta, Strunk,
  Sturani, Stuver, Sudhagar, Sudhir, Suh, Summerscales, Sun, Sun, Sunil, Sur,
  Suresh, Sutton, Swinkels, Szczepańczyk, Szewczyk, Tacca, Tait, Talbot,
  Talbot, Tanasijczuk, Tanner, Tao, Tao, Martín, Taranto, Tasson, Tenorio,
  Terhune, Terkowski, Thirugnanasambandam, Thomas, Thomas, Thomas, Thompson,
  Thondapu, Thorne, Thrane, Tiwari, Tiwari, Tiwari, Toivonen, Toland, Tolley,
  Tonelli, Torres-Forné, Torrie, e~Melo, Töyrä, Trapananti, Travasso,
  Traylor, Trevor, Tringali, Tripathee, Troiano, Trovato, Trozzo, Trudeau,
  Tsai, Tsai, Tsang, Tse, Tso, Tsukada, Tsuna, Tsutsui, Turbang, Turconi, Ubhi,
  Udall, Ueno, Unnikrishnan, Urban, Utina, Vahlbruch, Vajente, Vajpeyi, Valdes,
  Valentini, Valsan, van Bakel, van Beuzekom, van~den Brand, Broeck,
  Vander-Hyde, van~der Schaaf, van Heijningen, Vanosky, van Remortel, Vardaro,
  Vargas, Varma, Vasúth, Vecchio, Vedovato, Veitch, Veitch, Venneberg,
  Venugopalan, Verkindt, Verma, Verma, Veske, Vetrano, Viceré, Vidyant, Viets,
  Vijaykumar, Villa-Ortega, Vinet, Virtuoso, Vitale, Vo, Vocca, von Reis, von
  Wrangel, Vorvick, Vyatchanin, Wade, Wade, Wagner, Walet, Walker, Wallace,
  Wallace, Walsh, Wang, Wang, Ward, Warner, Was, Washington, Watchi, Weaver,
  Webster, Weinert, Weinstein, Weiss, Weller, Weller, Wellmann, Wen, Weßels,
  Wette, Whelan, White, Whiting, Whittle, Wilken, Williams, Williams,
  Williamson, Willis, Willke, Wilson, Winkler, Wipf, Wlodarczyk, Woan, Woehler,
  Wofford, Wong, Wu, Wysocki, Xiao, Yamamoto, Yang, Yang, Yang, Yang, Yap,
  Yeeles, Yelikar, Ying, Yoo, Yu, Yu, Zadrożny, Zanolin, Zelenova, Zendri,
  Zevin, Zhang, Zhang, Zhang, Zhang, Zhao, Zhao, Zhao, Zhou, Zhou, Zhu,
  Zimmerman, Zlochower, Zucker, \& Zweizig}]{LIGOScientific:2021usb}
Collaboration, T. L.~S., the Virgo~Collaboration, Abbott, R., {et~al.} 2021,
  arXiv:2108.01045

\bibitem[{Coughlin {et~al.}(2019{\natexlab{a}})Coughlin, Dietrich, Margalit, \&
  Metzger}]{Coughlin:2018fis}
Coughlin, M.~W., Dietrich, T., Margalit, B., \& Metzger, B.~D.
  2019{\natexlab{a}}, Mon. Not. Roy. Astron. Soc., 489, L91

\bibitem[{Coughlin {et~al.}(2018{\natexlab{a}})Coughlin, Dietrich, Doctor,
  Kasen, Coughlin, Jerkstrand, Leloudas, McBrien, Metzger, O’Shaughnessy, \&
  et~al.}]{Coughlin:2018miv}
Coughlin, M.~W., Dietrich, T., Doctor, Z., {et~al.} 2018{\natexlab{a}}, Monthly
  Notices of the Royal Astronomical Society, 480, 3871–3878.
\newblock \url{http://dx.doi.org/10.1093/mnras/sty2174}

\bibitem[{Coughlin {et~al.}(2018{\natexlab{b}})Coughlin, Tao, Chan, Chatterjee,
  Christensen, Ghosh, Greco, Hu, Kapadia, Rana, \& et~al.}]{Coughlin:2018lta}
Coughlin, M.~W., Tao, D., Chan, M.~L., {et~al.} 2018{\natexlab{b}}, Monthly
  Notices of the Royal Astronomical Society, 478, 692–702.
\newblock \url{http://dx.doi.org/10.1093/mnras/sty1066}

\bibitem[{Coughlin {et~al.}(2019{\natexlab{b}})Coughlin, Antier, Corre,
  Alqassimi, Anand, Christensen, Coulter, Foley, Guessoum, Mikulski, \&
  et~al.}]{Coughlin:2019qkn}
Coughlin, M.~W., Antier, S., Corre, D., {et~al.} 2019{\natexlab{b}}, Monthly
  Notices of the Royal Astronomical Society, 489, 5775–5783.
\newblock \url{http://dx.doi.org/10.1093/mnras/stz2485}

\bibitem[{Coughlin {et~al.}(2019{\natexlab{c}})Coughlin, Ahumada, Cenko,
  Cunningham, Ghosh, Singer, Bellm, Burns, De, Goldstein, \&
  et~al.}]{Coughlin:2019qov}
Coughlin, M.~W., Ahumada, T., Cenko, S.~B., {et~al.} 2019{\natexlab{c}},
  Publications of the Astronomical Society of the Pacific, 131, 048001.
\newblock \url{http://dx.doi.org/10.1088/1538-3873/aaff99}

\bibitem[{Coughlin {et~al.}(2020{\natexlab{a}})Coughlin, Dietrich, Heinzel,
  Khetan, Antier, Bulla, Christensen, Coulter, \& Foley}]{Coughlin:2019vtv}
Coughlin, M.~W., Dietrich, T., Heinzel, J., {et~al.} 2020{\natexlab{a}}, Phys.
  Rev. Res., 2, 022006

\bibitem[{Coughlin {et~al.}(2020{\natexlab{b}})Coughlin, Dietrich, Antier,
  Almualla, Anand, Bulla, Foucart, Guessoum, Hotokezaka, Kumar, \&
  et~al.}]{Coughlin:2020fwx}
Coughlin, M.~W., Dietrich, T., Antier, S., {et~al.} 2020{\natexlab{b}}, Monthly
  Notices of the Royal Astronomical Society, 497, 1181–1196.
\newblock \url{http://dx.doi.org/10.1093/mnras/staa1925}

\bibitem[{Coughlin {et~al.}(2020{\natexlab{c}})Coughlin, Dietrich, Antier,
  Bulla, Foucart, Hotokezaka, Raaijmakers, Hinderer, \&
  Nissanke}]{Coughlin:2019zqi}
---. 2020{\natexlab{c}}, Mon. Not. Roy. Astron. Soc., 492, 863

\bibitem[{Coulter {et~al.}(2017)Coulter, Foley, Kilpatrick, Drout, Piro,
  Shappee, Siebert, Simon, Ulloa, Kasen, \& et~al.}]{Coulter:2017wya}
Coulter, D.~A., Foley, R.~J., Kilpatrick, C.~D., {et~al.} 2017, Science, 358,
  1556–1558.
\newblock \url{http://dx.doi.org/10.1126/science.aap9811}

\bibitem[{Cowperthwaite \& Berger(2015)}]{Cowperthwaite:2015kya}
Cowperthwaite, P.~S., \& Berger, E. 2015, Astrophys. J., 814, 25

\bibitem[{Cutler {et~al.}(1993)Cutler, Apostolatos, Bildsten, Finn, Flanagan,
  Kennefick, Markovic, Ori, Poisson, Sussman, \& et~al.}]{Cutler:1992tc}
Cutler, C., Apostolatos, T.~A., Bildsten, L., {et~al.} 1993, Physical Review
  Letters, 70, 2984–2987.
\newblock \url{http://dx.doi.org/10.1103/PhysRevLett.70.2984}

\bibitem[{Dal~Canton {et~al.}(2020)Dal~Canton, Nitz, Gadre, Davies,
  Villa-Ortega, Dent, Harry, \& Xiao}]{DalCanton:2020vpm}
Dal~Canton, T., Nitz, A.~H., Gadre, B., {et~al.} 2020, arXiv:2008.07494

\bibitem[{De {et~al.}(2020)De, Hankins, Kasliwal, Moore, Ofek, Adams, Ashley,
  Babul, Bagdasaryan, Burdge, \& et~al.}]{De:2019xhw}
De, K., Hankins, M.~J., Kasliwal, M.~M., {et~al.} 2020, Publications of the
  Astronomical Society of the Pacific, 132, 025001.
\newblock \url{http://dx.doi.org/10.1088/1538-3873/ab6069}

\bibitem[{Dichiara {et~al.}(2021)Dichiara, Becerra, Chase, Troja, Lee, Watson,
  Butler, O’Connor, Pereyra, López, \& et~al.}]{Dichiara:2021vjy}
Dichiara, S., Becerra, R.~L., Chase, E.~A., {et~al.} 2021, The Astrophysical
  Journal Letters, 923, L32.
\newblock \url{http://dx.doi.org/10.3847/2041-8213/ac4259}

\bibitem[{Dietrich {et~al.}(2020)Dietrich, Coughlin, Pang, Bulla, Heinzel,
  Issa, Tews, \& Antier}]{Dietrich:2020efo}
Dietrich, T., Coughlin, M.~W., Pang, P. T.~H., {et~al.} 2020, Science, 370,
  1450

\bibitem[{Dobie {et~al.}(2019)Dobie, Stewart, Murphy, Lenc, Wang, Kaplan,
  Andreoni, Banfield, Brown, Corsi, \& et~al.}]{Dobie:2019ctw}
Dobie, D., Stewart, A., Murphy, T., {et~al.} 2019, The Astrophysical Journal,
  887, L13.
\newblock \url{http://dx.doi.org/10.3847/2041-8213/ab59db}

\bibitem[{Evans {et~al.}(2017)Evans, Cenko, Kennea, Emery, Kuin, Korobkin,
  Wollaeger, Fryer, Madsen, Harrison, \& et~al.}]{Evans:2017mmy}
Evans, P.~A., Cenko, S.~B., Kennea, J.~A., {et~al.} 2017, Science, 358,
  1565–1570.
\newblock \url{http://dx.doi.org/10.1126/science.aap9580}

\bibitem[{Farrow {et~al.}(2019)Farrow, Zhu, \& Thrane}]{Farrow:2019xnc}
Farrow, N., Zhu, X.-J., \& Thrane, E. 2019, Astrophys. J., 876, 18

\bibitem[{Fern\'andez {et~al.}(2017)Fern\'andez, Foucart, Kasen, Lippuner,
  Desai, \& Roberts}]{Fernandez:2016sbf}
Fern\'andez, R., Foucart, F., Kasen, D., {et~al.} 2017, Class. Quant. Grav.,
  34, 154001

\bibitem[{Feroz \& Hobson(2008)}]{Feroz:2007kg}
Feroz, F., \& Hobson, M.~P. 2008, Mon. Not. Roy. Astron. Soc., 384, 449

\bibitem[{Feroz {et~al.}(2009)Feroz, Hobson, \& Bridges}]{Feroz:2008xx}
Feroz, F., Hobson, M.~P., \& Bridges, M. 2009, Mon. Not. Roy. Astron. Soc.,
  398, 1601

\bibitem[{Feroz {et~al.}(2019)Feroz, Hobson, Cameron, \&
  Pettitt}]{Feroz:2013hea}
Feroz, F., Hobson, M.~P., Cameron, E., \& Pettitt, A.~N. 2019, Open J.
  Astrophys., 2, 10

\bibitem[{Finstad \& Brown(2020)}]{Finstad:2020sok}
Finstad, D., \& Brown, D.~A. 2020, Astrophys. J. Lett., 905, L9

\bibitem[{Fong {et~al.}(2021)Fong, Laskar, Rastinejad, Escorial, Schroeder,
  Barnes, Kilpatrick, Paterson, Berger, Metzger, \& et~al.}]{Fong:2020cej}
Fong, W., Laskar, T., Rastinejad, J., {et~al.} 2021, The Astrophysical Journal,
  906, 127.
\newblock \url{http://dx.doi.org/10.3847/1538-4357/abc74a}

\bibitem[{{Frostig} {et~al.}(2020){Frostig}, {Baker}, {Brown}, {Burruss},
  {Clark}, {F{\.z}r{\'e}sz}, {Ganciu}, {Hinrichsen}, {Karambelkar}, {Kasliwal},
  {Lourie}, {Malonis}, {Simcoe}, \& {Zolkower}}]{Frostig:2020}
{Frostig}, D., {Baker}, J.~W., {Brown}, J., {et~al.} 2020, in Society of
  Photo-Optical Instrumentation Engineers (SPIE) Conference Series, Vol. 11447,
  Society of Photo-Optical Instrumentation Engineers (SPIE) Conference Series,
  1144767

\bibitem[{Frostig {et~al.}(2021)Frostig, Biscoveanu, Mo, Karambelkar,
  Dal~Canton, Chen, Kasliwal, Katsavounidis, Lourie, Simcoe, \&
  Vitale}]{frostig_danielle_2021_5507322}
Frostig, D., Biscoveanu, S., Mo, G., {et~al.} 2021, {Public release of data
  associated with WINTER/LIGO BNS kilonova simulations}, vv1,  Zenodo,
  doi:10.5281/zenodo.5507322.
\newblock \url{https://doi.org/10.5281/zenodo.5507322}

\bibitem[{Förster {et~al.}(2021)Förster, Cabrera-Vives, Castillo-Navarrete,
  Estévez, Sánchez-Sáez, Arredondo, Bauer, Carrasco-Davis, Catelan,
  Elorrieta, \& et~al.}]{Forster:2020pxc}
Förster, F., Cabrera-Vives, G., Castillo-Navarrete, E., {et~al.} 2021, The
  Astronomical Journal, 161, 242.
\newblock \url{http://dx.doi.org/10.3847/1538-3881/abe9bc}

\bibitem[{Gabbard {et~al.}(2021)Gabbard, Messenger, Heng, Tonolini, \&
  Murray-Smith}]{Gabbard:2019rde}
Gabbard, H., Messenger, C., Heng, I.~S., Tonolini, F., \& Murray-Smith, R.
  2021, Nature Physics, doi:10.1038/s41567-021-01425-7.
\newblock \url{http://dx.doi.org/10.1038/s41567-021-01425-7}

\bibitem[{Ghosh {et~al.}(2016)Ghosh, Bloemen, Nelemans, Groot, \&
  Price}]{Ghosh:2015sxp}
Ghosh, S., Bloemen, S., Nelemans, G., Groot, P.~J., \& Price, L.~R. 2016,
  Astron. Astrophys., 592, A82

\bibitem[{Goldstein {et~al.}(2017)Goldstein, Veres, Burns, Briggs, Hamburg,
  Kocevski, Wilson-Hodge, Preece, Poolakkil, Roberts, \&
  et~al.}]{Goldstein:2017mmi}
Goldstein, A., Veres, P., Burns, E., {et~al.} 2017, The Astrophysical Journal,
  848, L14.
\newblock \url{http://dx.doi.org/10.3847/2041-8213/aa8f41}

\bibitem[{Goldstein {et~al.}(2015)Goldstein, D’Andrea, Fischer, Foley, Gupta,
  Kessler, Kim, Nichol, Nugent, Papadopoulos, \& et~al.}]{DES:2015sji}
Goldstein, D.~A., D’Andrea, C.~B., Fischer, J.~A., {et~al.} 2015, The
  Astronomical Journal, 150, 82.
\newblock \url{http://dx.doi.org/10.1088/0004-6256/150/3/82}

\bibitem[{Gompertz {et~al.}(2020)Gompertz, Cutter, Steeghs, Galloway, Lyman,
  Ulaczyk, Dyer, Ackley, Dhillon, O’Brien, \& et~al.}]{Gompertz:2020cur}
Gompertz, B.~P., Cutter, R., Steeghs, D., {et~al.} 2020, Monthly Notices of the
  Royal Astronomical Society, 497, 726–738.
\newblock \url{http://dx.doi.org/10.1093/mnras/staa1845}

\bibitem[{{Green} \& {Gair}(2020)}]{Green:2020dnx}
{Green}, S.~R., \& {Gair}, J. 2020, arXiv e-prints, arXiv:2008.03312

\bibitem[{Green {et~al.}(2020)Green, Simpson, \& Gair}]{Green:2020hst}
Green, S.~R., Simpson, C., \& Gair, J. 2020, Phys. Rev. D, 102, 104057

\bibitem[{Grossman {et~al.}(2014)Grossman, Korobkin, Rosswog, \&
  Piran}]{Grossman:2013lqa}
Grossman, D., Korobkin, O., Rosswog, S., \& Piran, T. 2014, Mon. Not. Roy.
  Astron. Soc., 439, 757

\bibitem[{Gómez {et~al.}(2020)Gómez, Neira, Hernández Hoyos, Arbeláez, \&
  Forero-Romero}]{G_mez_2020}
Gómez, C., Neira, M., Hernández Hoyos, M., Arbeláez, P., \& Forero-Romero,
  J.~E. 2020, Monthly Notices of the Royal Astronomical Society, 499,
  3130–3138.
\newblock \url{http://dx.doi.org/10.1093/mnras/staa2973}

\bibitem[{Haggard {et~al.}(2017)Haggard, Nynka, Ruan, Kalogera, Bradley~Cenko,
  Evans, \& Kennea}]{Haggard:2017qne}
Haggard, D., Nynka, M., Ruan, J.~J., {et~al.} 2017, Astrophys. J. Lett., 848,
  L25

\bibitem[{Hallinan {et~al.}(2017)Hallinan, Corsi, Mooley, Hotokezaka, Nakar,
  Kasliwal, Kaplan, Frail, Myers, Murphy, \& et~al.}]{Hallinan:2017woc}
Hallinan, G., Corsi, A., Mooley, K.~P., {et~al.} 2017, Science, 358,
  1579–1583.
\newblock \url{http://dx.doi.org/10.1126/science.aap9855}

\bibitem[{Hannam {et~al.}(2014)Hannam, Schmidt, Boh\'e, Haegel, Husa, Ohme,
  Pratten, \& P\"urrer}]{Hannam:2013oca}
Hannam, M., Schmidt, P., Boh\'e, A., {et~al.} 2014, Phys. Rev. Lett., 113,
  151101

\bibitem[{Harris {et~al.}(2020)Harris, Millman, van~der Walt, Gommers,
  Virtanen, Cournapeau, Wieser, Taylor, Berg, Smith, Kern, Picus, Hoyer, van
  Kerkwijk, Brett, Haldane, del R{\'{i}}o, Wiebe, Peterson,
  G{\'{e}}rard-Marchant, Sheppard, Reddy, Weckesser, Abbasi, Gohlke, \&
  Oliphant}]{harris2020array}
Harris, C.~R., Millman, K.~J., van~der Walt, S.~J., {et~al.} 2020, Nature, 585,
  357.
\newblock \url{https://doi.org/10.1038/s41586-020-2649-2}

\bibitem[{Harry {et~al.}(2009)Harry, Allen, \& Sathyaprakash}]{Harry:2009ea}
Harry, I.~W., Allen, B., \& Sathyaprakash, B.~S. 2009, Phys. Rev. D, 80, 104014

\bibitem[{Heuvel(2017)}]{Heuvel:2017ziq}
Heuvel, E. P. J. v.~d. 2017, J. Astrophys. Astron., 38, 45

\bibitem[{Holz \& Hughes(2005)}]{Holz:2005df}
Holz, D.~E., \& Hughes, S.~A. 2005, Astrophys. J., 629, 15

\bibitem[{{Hu} {et~al.}(2021){Hu}, {Li}, {Castro-Tirado},
  {Fernandez-Garc{\'\i}a}, {Castell{\'o}n}, {Carrasco-Garc{\'\i}a}, {Perez del
  Pulgar}, {Caballero-Garc{\'\i}a}, {Querel}, {Bai}, {Fan}, {Guziy}, {Wang},
  {Xiong}, {Xin}, {Zhao}, {Hiriart}, {Lee}, {Jeong}, \&
  {Park}}]{2021RMxAC..53...75H}
{Hu}, Y.~D., {Li}, X.~Y., {Castro-Tirado}, A.~J., {et~al.} 2021, in Revista
  Mexicana de Astronomia y Astrofisica Conference Series, Vol.~53, Revista
  Mexicana de Astronomia y Astrofisica Conference Series, 75--82

\bibitem[{Hunter(2007)}]{Hunter:2007}
Hunter, J.~D. 2007, Matplotlib: A 2D graphics environment,  IEEE COMPUTER SOC,
  doi:10.1109/MCSE.2007.55

\bibitem[{Husa {et~al.}(2016)Husa, Khan, Hannam, P\"urrer, Ohme,
  Jim\'enez~Forteza, \& Boh\'e}]{Husa:2015iqa}
Husa, S., Khan, S., Hannam, M., {et~al.} 2016, Phys. Rev. D, 93, 044006

\bibitem[{Kasen {et~al.}(2013)Kasen, Badnell, \& Barnes}]{Kasen:2013xka}
Kasen, D., Badnell, N.~R., \& Barnes, J. 2013, Astrophys. J., 774, 25

\bibitem[{Kasen {et~al.}(2017)Kasen, Metzger, Barnes, Quataert, \&
  Ramirez-Ruiz}]{Kasen:2017sxr}
Kasen, D., Metzger, B., Barnes, J., Quataert, E., \& Ramirez-Ruiz, E. 2017,
  Nature, 551, 80

\bibitem[{Kasliwal {et~al.}(2019{\natexlab{a}})Kasliwal, Kasen, Lau, Perley,
  Rosswog, Ofek, Hotokezaka, Chary, Sollerman, Goobar, \&
  et~al.}]{Kasliwal:2018fwk}
Kasliwal, M.~M., Kasen, D., Lau, R.~M., {et~al.} 2019{\natexlab{a}}, Monthly
  Notices of the Royal Astronomical Society: Letters, 510, L7–L12.
\newblock \url{http://dx.doi.org/10.1093/mnrasl/slz007}

\bibitem[{Kasliwal {et~al.}(2019{\natexlab{b}})Kasliwal, Cannella, Bagdasaryan,
  Hung, Feindt, Singer, Coughlin, Fremling, Walters, Duev, \&
  et~al.}]{Kasliwal:2019pyx}
Kasliwal, M.~M., Cannella, C., Bagdasaryan, A., {et~al.} 2019{\natexlab{b}},
  Publications of the Astronomical Society of the Pacific, 131, 038003.
\newblock \url{http://dx.doi.org/10.1088/1538-3873/aafbc2}

\bibitem[{Kasliwal {et~al.}(2020)Kasliwal, Anand, Ahumada, Stein, Carracedo,
  Andreoni, Coughlin, Singer, Kool, De, \& et~al.}]{Kasliwal:2020wmy}
Kasliwal, M.~M., Anand, S., Ahumada, T., {et~al.} 2020, The Astrophysical
  Journal, 905, 145.
\newblock \url{http://dx.doi.org/10.3847/1538-4357/abc335}

\bibitem[{Khan {et~al.}(2016)Khan, Husa, Hannam, Ohme, P\"urrer,
  Jim\'enez~Forteza, \& Boh\'e}]{Khan:2015jqa}
Khan, S., Husa, S., Hannam, M., {et~al.} 2016, Phys. Rev. D, 93, 044007

\bibitem[{Kochanek {et~al.}(2017)Kochanek, Shappee, Stanek, Holoien, Thompson,
  Prieto, Dong, Shields, Will, Britt, \& et~al.}]{Kochanek:2017wud}
Kochanek, C.~S., Shappee, B.~J., Stanek, K.~Z., {et~al.} 2017, Publications of
  the Astronomical Society of the Pacific, 129, 104502.
\newblock \url{http://dx.doi.org/10.1088/1538-3873/aa80d9}

\bibitem[{Lawrence {et~al.}(2007)Lawrence, Warren, Almaini, Edge, Hambly,
  Jameson, Lucas, Casali, Adamson, Dye, \& et~al.}]{Lawrence:2006de}
Lawrence, A., Warren, S.~J., Almaini, O., {et~al.} 2007, Monthly Notices of the
  Royal Astronomical Society, 379, 1599–1617.
\newblock \url{http://dx.doi.org/10.1111/j.1365-2966.2007.12040.x}

\bibitem[{Levan(2020)}]{Levan:2020BP}
Levan, A. 2020, PoS, Asterics2019, 044

\bibitem[{Li \& Paczynski(1998)}]{Li:1998bw}
Li, L.-X., \& Paczynski, B. 1998, Astrophys. J. Lett., 507, L59

\bibitem[{Lipunov {et~al.}(2017)Lipunov, Gorbovskoy, Kornilov, Tyurina,
  Balanutsa, Kuznetsov, Vlasenko, Kuvshinov, Gorbunov, Buckley, \&
  et~al.}]{Lipunov:2017dwd}
Lipunov, V.~M., Gorbovskoy, E., Kornilov, V.~G., {et~al.} 2017, The
  Astrophysical Journal, 850, L1.
\newblock \url{http://dx.doi.org/10.3847/2041-8213/aa92c0}

\bibitem[{Lorimer(2008)}]{Lorimer:2008se}
Lorimer, D.~R. 2008, Living Rev. Rel., 11, 8

\bibitem[{Lourie {et~al.}(2020)Lourie, Baker, Burruss, Egan, Fűrész, Frostig,
  Garcia-Zych, Ganciu, Haworth, Hinrichsen, Kasliwal, Karambelkar, Malonis,
  Simcoe, \& Zolkower}]{Lourie:2020}
Lourie, N.~P., Baker, J.~W., Burruss, R.~S., {et~al.} 2020, in Ground-based and
  Airborne Instrumentation for Astronomy VIII, ed. C.~J. Evans, J.~J. Bryant,
  \& K.~Motohara, Vol. 11447, International Society for Optics and Photonics
  (SPIE), 2064 -- 2077.
\newblock \url{https://doi.org/10.1117/12.2561210}

\bibitem[{Lyne {et~al.}(2004)Lyne, Burgay, Kramer, Possenti, Manchester,
  Camilo, McLaughlin, Lorimer, D’Amico, Joshi, \& et~al.}]{Lyne:2004cj}
Lyne, A.~G., Burgay, M., Kramer, M., {et~al.} 2004, Science, 303, 1153–1157.
\newblock \url{http://dx.doi.org/10.1126/science.1094645}

\bibitem[{Magee {et~al.}(2021)Magee, Chatterjee, Singer, Sachdev, Kovalam, Mo,
  Anderson, Brady, Brockill, Cannon, \& et~al.}]{Magee:2021xdx}
Magee, R., Chatterjee, D., Singer, L.~P., {et~al.} 2021, The Astrophysical
  Journal Letters, 910, L21.
\newblock \url{http://dx.doi.org/10.3847/2041-8213/abed54}

\bibitem[{Mahabal {et~al.}(2019)Mahabal, Rebbapragada, Walters, Masci,
  Blagorodnova, Roestel, Ye, Biswas, Burdge, Chang, \& et~al.}]{Mahabal_2019}
Mahabal, A., Rebbapragada, U., Walters, R., {et~al.} 2019, Publications of the
  Astronomical Society of the Pacific, 131, 038002.
\newblock \url{http://dx.doi.org/10.1088/1538-3873/aaf3fa}

\bibitem[{Malonis {et~al.}(2020)Malonis, Lourie, Fűrész, Frostig, Hinrichsen,
  \& Simcoe}]{Malonis:2020}
Malonis, A.~C., Lourie, N.~P., Fűrész, G., {et~al.} 2020, in X-Ray, Optical,
  and Infrared Detectors for Astronomy IX, ed. A.~D. Holland \& J.~Beletic,
  Vol. 11454, International Society for Optics and Photonics (SPIE), 514 --
  519.
\newblock \url{https://doi.org/10.1117/12.2561228}

\bibitem[{Margalit \& Metzger(2017)}]{Margalit:2017dij}
Margalit, B., \& Metzger, B.~D. 2017, Astrophys. J. Lett., 850, L19

\bibitem[{Margutti {et~al.}(2017)Margutti, Berger, Fong, Guidorzi, Alexander,
  Metzger, Blanchard, Cowperthwaite, Chornock, Eftekhari, \&
  et~al.}]{Margutti:2017cjl}
Margutti, R., Berger, E., Fong, W., {et~al.} 2017, The Astrophysical Journal,
  848, L20.
\newblock \url{http://dx.doi.org/10.3847/2041-8213/aa9057}

\bibitem[{McKinney(2010)}]{mckinney2010data}
McKinney, W. 2010, Data structures for statistical computing in python,
  Proceedings of the 9th Python in Science Conference

\bibitem[{Metzger(2020)}]{Metzger:2019zeh}
Metzger, B.~D. 2020, Living Rev. Rel., 23, 1

\bibitem[{Metzger {et~al.}(2010)Metzger, Martinez-Pinedo, Darbha, Quataert,
  Arcones, Kasen, Thomas, Nugent, Panov, \& Zinner}]{Metzger:2010sy}
Metzger, B.~D., Martinez-Pinedo, G., Darbha, S., {et~al.} 2010, Mon. Not. Roy.
  Astron. Soc., 406, 2650

\bibitem[{Morisaki \& Raymond(2020)}]{Morisaki:2020oqk}
Morisaki, S., \& Raymond, V. 2020, Phys. Rev. D, 102, 104020

\bibitem[{Nitz {et~al.}(2018)Nitz, Dal~Canton, Davis, \& Reyes}]{Nitz:2018rgo}
Nitz, A.~H., Dal~Canton, T., Davis, D., \& Reyes, S. 2018, Phys. Rev. D, 98,
  024050

\bibitem[{Oates {et~al.}(2021)Oates, Marshall, Breeveld, Kuin, Brown,
  De~Pasquale, Evans, Fenney, Gronwall, Kennea, \& et~al.}]{Oates:2021eyk}
Oates, S.~R., Marshall, F.~E., Breeveld, A.~A., {et~al.} 2021, Monthly Notices
  of the Royal Astronomical Society, 507, 1296–1317.
\newblock \url{http://dx.doi.org/10.1093/mnras/stab2189}

\bibitem[{Page {et~al.}(2020)Page, Evans, Tohuvavohu, Kennea, Klingler, Cenko,
  Oates, Ambrosi, Barthelmy, Beardmore, \& et~al.}]{Page:2020tnx}
Page, K.~L., Evans, P.~A., Tohuvavohu, A., {et~al.} 2020, Monthly Notices of
  the Royal Astronomical Society, 499, 3459–3480.
\newblock \url{http://dx.doi.org/10.1093/mnras/staa3032}

\bibitem[{Pankow {et~al.}(2015)Pankow, Brady, Ochsner, \&
  O'Shaughnessy}]{Pankow:2015cra}
Pankow, C., Brady, P., Ochsner, E., \& O'Shaughnessy, R. 2015, Phys. Rev. D,
  92, 023002

\bibitem[{Paterson {et~al.}(2021)Paterson, Lundquist, Rastinejad, Fong, Sand,
  Andrews, Amaro, Eskandari, Wyatt, Daly, \& et~al.}]{Paterson:2020mmd}
Paterson, K., Lundquist, M.~J., Rastinejad, J.~C., {et~al.} 2021, The
  Astrophysical Journal, 912, 128.
\newblock \url{http://dx.doi.org/10.3847/1538-4357/abeb71}

\bibitem[{Petrov {et~al.}(2021)Petrov, Singer, Coughlin, Kumar, Almualla,
  Anand, Bulla, Dietrich, Foucart, \& Guessoum}]{Petrov:2021bqm}
Petrov, P., Singer, L.~P., Coughlin, M.~W., {et~al.} 2021, arXiv:2108.07277

\bibitem[{Pian {et~al.}(2017)Pian, D’Avanzo, Benetti, Branchesi, Brocato,
  Campana, Cappellaro, Covino, D’Elia, Fynbo, \& et~al.}]{Pian:2017gtc}
Pian, E., D’Avanzo, P., Benetti, S., {et~al.} 2017, Nature, 551, 67–70.
\newblock \url{http://dx.doi.org/10.1038/nature24298}

\bibitem[{Privitera {et~al.}(2014)Privitera, Mohapatra, Ajith, Cannon,
  Fotopoulos, Frei, Hanna, Weinstein, \& Whelan}]{Privitera:2013xza}
Privitera, S., Mohapatra, S. R.~P., Ajith, P., {et~al.} 2014, Phys. Rev. D, 89,
  024003

\bibitem[{{Radhakrishnan} \& {Srinivasan}(1982)}]{1982CSci...51.1096R}
{Radhakrishnan}, V., \& {Srinivasan}, G. 1982, Current Science, 51, 1096

\bibitem[{Roberts {et~al.}(2011)Roberts, Kasen, Lee, \&
  Ramirez-Ruiz}]{Roberts:2011xz}
Roberts, L.~F., Kasen, D., Lee, W.~H., \& Ramirez-Ruiz, E. 2011, Astrophys. J.
  Lett., 736, L21

\bibitem[{Robitaille {et~al.}(2013)Robitaille, Tollerud, Greenfield,
  Droettboom, Bray, Aldcroft, Davis, Ginsburg, Price-Whelan, \&
  et~al.}]{astropy:2013}
Robitaille, T.~P., Tollerud, E.~J., Greenfield, P., {et~al.} 2013, Astropy: A
  community Python package for astronomy,  EDP Sciences,
  doi:10.1051/0004-6361/201322068.
\newblock \url{http://dx.doi.org/10.1051/0004-6361/201322068}

\bibitem[{Romano \& Cornish(2017)}]{Romano:2016dpx}
Romano, J.~D., \& Cornish, N.~J. 2017, Living Rev. Rel., 20, 2

\bibitem[{Romero-Shaw {et~al.}(2020)Romero-Shaw, Talbot, Biscoveanu,
  D’Emilio, Ashton, Berry, Coughlin, Galaudage, Hoy, Hübner, \&
  et~al.}]{Romero-Shaw:2020owr}
Romero-Shaw, I.~M., Talbot, C., Biscoveanu, S., {et~al.} 2020, Monthly Notices
  of the Royal Astronomical Society, 499, 3295–3319.
\newblock \url{http://dx.doi.org/10.1093/mnras/staa2850}

\bibitem[{Rosswog(2005)}]{Rosswog:2005su}
Rosswog, S. 2005, Astrophys. J., 634, 1202

\bibitem[{Simcoe {et~al.}(2019)Simcoe, F\'{u}r\'esz, Sullivan, Hellickson,
  Malonis, Kasliwal, Shectman, Kollmeier, \& Moore}]{Simcoe:2019aps}
Simcoe, R.~A., F\'{u}r\'esz, G., Sullivan, P.~W., {et~al.} 2019, Astron. J.,
  157, 46

\bibitem[{Singer \& Price(2016)}]{Singer:2015ema}
Singer, L.~P., \& Price, L.~R. 2016, Phys. Rev. D, 93, 024013

\bibitem[{Singer {et~al.}(2014)Singer, Price, Farr, Urban, Pankow, Vitale,
  Veitch, Farr, Hanna, Cannon, \& et~al.}]{Singer:2014qca}
Singer, L.~P., Price, L.~R., Farr, B., {et~al.} 2014, The Astrophysical
  Journal, 795, 105.
\newblock \url{http://dx.doi.org/10.1088/0004-637X/795/2/105}

\bibitem[{Singer {et~al.}(2016{\natexlab{a}})Singer, Chen, Holz, Farr, Price,
  Raymond, Cenko, Gehrels, Cannizzo, Kasliwal, \& et~al.}]{Singer:2016eax}
Singer, L.~P., Chen, H.-Y., Holz, D.~E., {et~al.} 2016{\natexlab{a}}, The
  Astrophysical Journal, 829, L15.
\newblock \url{http://dx.doi.org/10.3847/2041-8205/829/1/L15}

\bibitem[{Singer {et~al.}(2016{\natexlab{b}})Singer, Chen, Holz, Farr, Price,
  Raymond, Cenko, Gehrels, Cannizzo, Kasliwal, \& et~al.}]{Singer:2016erz}
---. 2016{\natexlab{b}}, The Astrophysical Journal Supplement Series, 226, 10.
\newblock \url{http://dx.doi.org/10.3847/0067-0049/226/1/10}

\bibitem[{Smartt {et~al.}(2017)Smartt, Chen, Jerkstrand, Coughlin, Kankare,
  Sim, Fraser, Inserra, Maguire, Chambers, \& et~al.}]{Smartt:2017fuw}
Smartt, S.~J., Chen, T.-W., Jerkstrand, A., {et~al.} 2017, Nature, 551,
  75–79.
\newblock \url{http://dx.doi.org/10.1038/nature24303}

\bibitem[{Smith {et~al.}(2016)Smith, Field, Blackburn, Haster, P\"urrer,
  Raymond, \& Schmidt}]{Smith:2016qas}
Smith, R., Field, S.~E., Blackburn, K., {et~al.} 2016, Phys. Rev. D, 94, 044031

\bibitem[{Smith {et~al.}(2020)Smith, Ashton, Vajpeyi, \&
  Talbot}]{Smith:2019ucc}
Smith, R. J.~E., Ashton, G., Vajpeyi, A., \& Talbot, C. 2020, Mon. Not. Roy.
  Astron. Soc., 498, 4492

\bibitem[{Soares-Santos {et~al.}(2017)Soares-Santos, Holz, Annis, Chornock,
  Herner, Berger, Brout, Chen, Kessler, Sako, \& et~al.}]{DES:2017kbs}
Soares-Santos, M., Holz, D.~E., Annis, J., {et~al.} 2017, The Astrophysical
  Journal, 848, L16.
\newblock \url{http://dx.doi.org/10.3847/2041-8213/aa9059}

\bibitem[{Somiya(2012)}]{Somiya:2011np}
Somiya, K. 2012, Class. Quant. Grav., 29, 124007

\bibitem[{Sooknunan {et~al.}(2020)Sooknunan, Lochner, Bassett, Peiris, Fender,
  Stewart, Pietka, Woudt, McEwen, \& Lahav}]{Sooknunan_2020}
Sooknunan, K., Lochner, M., Bassett, B.~A., {et~al.} 2020, Monthly Notices of
  the Royal Astronomical Society, 502, 206–224.
\newblock \url{http://dx.doi.org/10.1093/mnras/staa3873}

\bibitem[{Stachie {et~al.}(2021)Stachie, Coughlin, Dietrich, Antier, Bulla,
  Christensen, Essick, Landry, Mours, Schianchi, \& et~al.}]{Stachie:2021noh}
Stachie, C., Coughlin, M.~W., Dietrich, T., {et~al.} 2021, Monthly Notices of
  the Royal Astronomical Society, 505, 4235–4248.
\newblock \url{http://dx.doi.org/10.1093/mnras/stab1492}

\bibitem[{Talbot {et~al.}(2019)Talbot, Smith, Thrane, \&
  Poole}]{Talbot:2019okv}
Talbot, C., Smith, R., Thrane, E., \& Poole, G.~B. 2019, Phys. Rev. D, 100,
  043030

\bibitem[{Tanaka \& Hotokezaka(2013)}]{Tanaka:2013ana}
Tanaka, M., \& Hotokezaka, K. 2013, Astrophys. J., 775, 113

\bibitem[{Tanvir {et~al.}(2013)Tanvir, Levan, Fruchter, Hjorth, Wiersema,
  Tunnicliffe, \& de~Ugarte~Postigo}]{Tanvir:2013pia}
Tanvir, N.~R., Levan, A.~J., Fruchter, A.~S., {et~al.} 2013, Nature, 500, 547

\bibitem[{Tauris {et~al.}(2017)Tauris, Kramer, Freire, Wex, Janka, Langer,
  Podsiadlowski, Bozzo, Chaty, Kruckow, \& et~al.}]{Tauris:2017omb}
Tauris, T.~M., Kramer, M., Freire, P. C.~C., {et~al.} 2017, The Astrophysical
  Journal, 846, 170.
\newblock \url{http://dx.doi.org/10.3847/1538-4357/aa7e89}

\bibitem[{{Thakur} {et~al.}(2020){Thakur}, {Dichiara}, {Troja}, {Chase},
  {S{\'a}nchez-Ram{\'\i}rez}, {Piro}, {Fryer}, {Butler}, {Watson}, {Wollaeger},
  {Ambrosi}, {Becerra Gonz{\'a}lez}, {Becerra}, {Bruni}, {Cenko}, {Cusumano},
  {D'A{\`\i}}, {Durbak}, {Fontes}, {Gatkine}, {Hungerford}, {Korobkin},
  {Kutyrev}, {Lee}, {Lotti}, {Minervini}, {Novara}, {Parola}, {Pereyra},
  {Ricci}, {Tiengo}, \& {Veilleux}}]{Thakur:2020yvu}
{Thakur}, A.~L., {Dichiara}, S., {Troja}, E., {et~al.} 2020, \mnras, 499, 3868

\bibitem[{{the LIGO Scientific Collaboration} {et~al.}(2021){the LIGO
  Scientific Collaboration}, {The Virgo Collaboration}, \& {The KAGRA
  Scientific Collaboration}}]{LIGOScientific:2021psn}
{the LIGO Scientific Collaboration}, {The Virgo Collaboration}, \& {The KAGRA
  Scientific Collaboration}. 2021, arXiv:2111.03634

\bibitem[{Thrane \& Talbot(2019)}]{Thrane:2018qnx}
Thrane, E., \& Talbot, C. 2019, Publ. Astron. Soc. Austral., 36, e010,
  [Erratum: Publ.Astron.Soc.Austral. 37, e036 (2020)]

\bibitem[{{Tonry} {et~al.}(2018){Tonry}, {Denneau}, {Heinze}, {Stalder},
  {Smith}, {Smartt}, {Stubbs}, {Weiland}, \& {Rest}}]{Tonry2018abc}
{Tonry}, J.~L., {Denneau}, L., {Heinze}, A.~N., {et~al.} 2018, \pasp, 130,
  064505

\bibitem[{Travouillon {et~al.}(2020)Travouillon, Moore, Soon, Galla, Adams,
  Hart, \& Taylor}]{DREAMS}
Travouillon, T.~D., Moore, A.~M., Soon, J., {et~al.} 2020, in Ground-based and
  Airborne Telescopes VIII, ed. H.~K. Marshall, J.~Spyromilio, \& T.~Usuda,
  Vol. 11445, International Society for Optics and Photonics (SPIE).
\newblock \url{https://doi.org/10.1117/12.2562980}

\bibitem[{Troja {et~al.}(2017)Troja, Piro, van Eerten, Wollaeger, Im, Fox,
  Butler, Cenko, Sakamoto, Fryer, \& et~al.}]{Troja:2017nqp}
Troja, E., Piro, L., van Eerten, H., {et~al.} 2017, Nature, 551, 71–74.
\newblock \url{http://dx.doi.org/10.1038/nature24290}

\bibitem[{Troja {et~al.}(2018)Troja, Ryan, Piro, van Eerten, Cenko, Yoon, Lee,
  Im, Sakamoto, Gatkine, \& et~al.}]{Troja:2018ybt}
Troja, E., Ryan, G., Piro, L., {et~al.} 2018, Nature Communications, 9,
  doi:10.1038/s41467-018-06558-7.
\newblock \url{http://dx.doi.org/10.1038/s41467-018-06558-7}

\bibitem[{Turpin {et~al.}(2020)Turpin, Ganet, Antier, Bertin, Xin, Leroy, Wu,
  Xu, Han, Cai, \& et~al.}]{Turpin:2020dti}
Turpin, D., Ganet, M., Antier, S., {et~al.} 2020, Monthly Notices of the Royal
  Astronomical Society, 497, 2641–2650.
\newblock \url{http://dx.doi.org/10.1093/mnras/staa2046}

\bibitem[{Veitch \& Vecchio(2010)}]{Veitch:2009hd}
Veitch, J., \& Vecchio, A. 2010, Phys. Rev. D, 81, 062003

\bibitem[{Veitch {et~al.}(2015)Veitch, Raymond, Farr, Farr, Graff, Vitale,
  Aylott, Blackburn, Christensen, Coughlin, \& et~al.}]{Veitch:2014wba}
Veitch, J., Raymond, V., Farr, B., {et~al.} 2015, Physical Review D, 91,
  doi:10.1103/physrevd.91.042003.
\newblock \url{http://dx.doi.org/10.1103/PhysRevD.91.042003}

\bibitem[{Williams {et~al.}(2021)Williams, Veitch, \&
  Messenger}]{Williams:2021qyt}
Williams, M.~J., Veitch, J., \& Messenger, C. 2021, Phys. Rev. D, 103, 103006

\bibitem[{Wollaeger {et~al.}(2021)Wollaeger, Fryer, Chase, Fontes, Ristic,
  Hungerford, Korobkin, O'Shaughnessy, \& Herring}]{Wollaeger:2021qgf}
Wollaeger, R.~T., Fryer, C.~L., Chase, E.~A., {et~al.} 2021, Astrophys. J.,
  918, 10

\bibitem[{Wysocki {et~al.}(2019)Wysocki, O'Shaughnessy, Lange, \&
  Fang}]{Wysocki:2019grj}
Wysocki, D., O'Shaughnessy, R., Lange, J., \& Fang, Y.-L.~L. 2019, Phys. Rev.
  D, 99, 084026

\bibitem[{You {et~al.}(2021)You, Ashton, Zhu, Thrane, \& Zhu}]{You:2021eeq}
You, Z.-Q., Ashton, G., Zhu, X.-J., Thrane, E., \& Zhu, Z.-H. 2021, Monthly
  Notices of the Royal Astronomical Society, 509, 3957–3965.
\newblock \url{http://dx.doi.org/10.1093/mnras/stab2977}

\bibitem[{Zackay {et~al.}(2018)Zackay, Dai, \& Venumadhav}]{Zackay:2018qdy}
Zackay, B., Dai, L., \& Venumadhav, T. 2018, arXiv:1806.08792

\bibitem[{Zhu {et~al.}(2021{\natexlab{a}})Zhu, Wu, Yang, Zhang, Yu, Gao, Cao,
  \& Liu}]{Zhu:2021ysz}
Zhu, J.-P., Wu, S., Yang, Y.-P., {et~al.} 2021{\natexlab{a}}, The Astrophysical
  Journal, 921, 156.
\newblock \url{http://dx.doi.org/10.3847/1538-4357/ac19a7}

\bibitem[{Zhu {et~al.}(2020)Zhu, Yang, Liu, Huang, Zhang, Li, Yu, \&
  Gao}]{Zhu:2020inc}
Zhu, J.-P., Yang, Y.-P., Liu, L.-D., {et~al.} 2020, Astrophys. J., 897, 20

\bibitem[{Zhu {et~al.}(2021{\natexlab{b}})Zhu, Wu, Yang, Zhang, Gao, Yu, Li,
  Cao, Liu, Huang, \& et~al.}]{Zhu:2020ffa}
Zhu, J.-P., Wu, S., Yang, Y.-P., {et~al.} 2021{\natexlab{b}}, The Astrophysical
  Journal, 917, 24.
\newblock \url{http://dx.doi.org/10.3847/1538-4357/abfe5e}

\end{thebibliography}
\bibliographystyle{aasjournal}

\end{document}